\documentclass[journal,11pt,draftclsnofoot,onecolumn]{IEEEtran}
\usepackage[noadjust]{cite}
\ifCLASSINFOpdf
\else
\usepackage[dvips]{graphicx,color}
\fi
\usepackage[cmex10]{amsmath}
\usepackage[T1]{fontenc}
\usepackage{ae}
\usepackage{aecompl}
\usepackage{amssymb}
\newtheorem{definition}{Definition}[section]
\newtheorem{theorem}{Theorem}[section]
\newtheorem{lemma}{Lemma}[section]
\newtheorem{remark}{Remark}[section]

\makeatletter
\renewcommand{\theequation}{\arabic{section}.\arabic{equation}}
\@addtoreset{equation}{section}
\makeatother
\makeatletter
    \renewcommand{\thefigure}{%
    \arabic{section}.\arabic{figure}}
\@addtoreset{figure}{section}
\makeatother
\begin{document}
%
% paper title
% can use linebreaks \\ within to get better formatting as desired
\title{Second-Order Resolvability, Intrinsic Randomness, and Fixed-Length Source Coding for Mixed Sources: Information Spectrum Approach}
%
%
% author names and IEEE memberships
% note positions of commas and nonbreaking spaces ( ~ ) LaTeX will not break
% a structure at a ~ so this keeps an author's name from being broken across
% two lines.
% use \thanks{} to gain access to the first footnote area
% a separate \thanks must be used for each paragraph as LaTeX2e's \thanks
% was not built to handle multiple paragraphs
%

\author{Ryo Nomura,~\IEEEmembership{Member,~IEEE,}
%        Toshiyasu~MATSUSHIMA,~\IEEEmembership{Member,~IEEE,}
        and~Te~Sun~Han,~\IEEEmembership{Life~Fellow,~IEEE}% <-this % stops a space
\thanks{R. Nomura is with the School of Network and Information, Senshu University, Kanagawa, Japan,
 e-mail: nomu@isc.senshu-u.ac.jp}% <-this % stops a space
\thanks{T. S. Han is with the National Institute of Information and Communications Technology (NICT), Tokyo, Japan, 
e-mail: han@is.uec.ac.jp }
%\thanks{T. MATSUSHIMA is with Waseda University.}% <-this % stops a space
%\thanks{Manuscript received April 19, 2005; revised January 11, 2007. }
%The material in this paper was presented in part \cite{Nomura_SITA2010} at the 33rd Symposium on Information Theory and its Applications, Nagano, Japan, December 2010.}
\thanks{The first author with this work was supported in part by JSPS Grant-in-Aid for Young Scientists (B) No. 23760346.}
}

\maketitle
\begin{abstract}
%\boldmath
The second-order achievable asymptotics in typical random number generation problems such as resolvability, intrinsic randomness, fixed-length source coding are considered. %resolvability problem is a kind of random number generation problems and formulated as follows. Given an arbitrary discrete random variable (called the target random number), we generate or approximate it by using a discrete uniform random number whose size is requested to be as small as possible.
In these problems, several researchers have derived the first-order and the second-order achievability rates for general sources using the information spectrum methods.
Although these formulas are general,  their computation are quite hard. 
Hence, an attempt to address explicit computation problems of achievable rates is meaningful. In particular, for i.i.d. sources, the second-order achievable rates have earlier been determined simply by using the asymptotic normality.
In this paper, we consider mixed sources of two i.i.d. sources. The mixed source is a typical case of nonergodic sources and whose self-information does not have the asymptotic normality. Nonetheless, we can explicitly compute the second-order achievable rates for these sources on the basis of two-peak asymptotic normality. 
In addition, extensions of our results to more general mixed sources, such as a mixture of countably infinite i.i.d. sources or Markovian sources, and a continuous mixture of i.i.d. sources, are considered.
\end{abstract}
% IEEEtran.cls defaults to using nonbold math in the Abstract.
% This preserves the distinction between vectors and scalars. However,
% if the journal you are submitting to favors bold math in the abstract,
% then you can use LaTeX's standard command \boldmath at the very start
% of the abstract to achieve this. Many IEEE journals frown on math
% in the abstract anyway.

% Note that keywords are not normally used for peerreview papers.

% For peer review papers, you can put extra information on the cover
% page as needed:
% \ifCLASSOPTIONpeerreview
% \begin{center} \bfseries EDICS Category: 3-BBND \end{center}
% \fi
%
% For peerreview papers, this IEEEtran command inserts a page break and
% creates the second title. It will be ignored for other modes.
\IEEEpeerreviewmaketitle

%\section{Introduction}
% The very first letter is a 2 line initial drop letter followed
% by the rest of the first word in caps.
% 
% form to use if the first word consists of a single letter:
% \IEEEPARstart{A}{demo} file is ....
% 
% form to use if you need the single drop letter followed by
% normal text (unknown if ever used by IEEE):
% \IEEEPARstart{A}{}demo file is ....
% 
% Some journals put the first two words in caps:
% \IEEEPARstart{T}{his demo} file is ....
% 
% Here we have the typical use of a "T" for an initial drop letter
% and "HIS" in caps to complete the first word.
%\IEEEPARstart{T}{his} demo file is intended to serve as a ``starter file''
%for IEEE journal papers produced under \LaTeX\ using
%IEEEtran.cls version 1.7 and later.
% You must have at least 2 lines in the paragraph with the drop letter
% (should never be an issue)

% needed in second column of first page if using \IEEEpubid
%\IEEEpubidadjcol

\section{Introduction}
\IEEEPARstart{T}{he} problem of random number generation is one of the main topics in information theory \cite{HV93,VV,Han_Folklore,VKV98}.
There are several problem settings in random number generation.
In particular, the resolvability problem and the intrinsic randomness problem are representative of them.  
The {\it resolvability} problem is formulated as follows \cite{Han}. 
We first use the term of \lq\lq {\it general source}" to denote a sequence ${\bf X} = \{ X^n \}_{n=1}^\infty$ of random variables $X^n$ indexed by $n$ (taking values in {\it countably infinite} sets), typically, $n$-dimensional random variables.
Given an arbitrary general source ${\bf X} = \{ X^n \}_{n=1}^\infty$ (called the {\it target} random number), we generate or approximate it by using a discrete {\it uniform} random number whose size is requested to be as small as possible.
% under the condition that the distance between the probability distribution of the source and the probability distribution of random variables to be generated is smaller than or equal to some small constant.
One of the main objectives in this problem is to construct an efficient algorithm that transforms the discrete uniform random number to the specified target source. %, and thereby to obtain the achievability condition. 
%The achievable condition means the condition for the source that there exists an algorithm under the condition that the distance between the probability distribution of the source and the probability distribution of random variables to be generated is smaller than or equal to some constant.
Han and Verd\'{u} \cite{HV93}, and Steinberg and Verd\'{u} \cite{Steinberg} have determined the infima of achievable uniform random number rates by using the information spectrum methods.
On the other hand, the {\it intrinsic randomness} problem is formulated as follows in Vembu and Verd\'{u} \cite{VV}: Given an arbitrary general source ${\bf X} = \{ X^n \}_{n=1}^\infty$ (called the {\it coin} source), we try to generate or approximate, by using ${\bf X} = \{ X^n \}_{n=1}^\infty$, a uniform random number at as large rates as possible. Vembu and Verd\'{u} \cite{VV} and Han \cite{Han} have determined the suprema of achievable uniform random number generation rates, again by invoking the information spectrum methods.
Since the class of {\it general} sources is quite large and it includes all nonstationary and/or nonergodic sources, their results are very basic and quite fundamental.

Furthermore, it turned out that these random number generation problems have close bearing with the fixed-length source coding problem (cf. Han and Verd\'{u} \cite{HV93}). All the formulas established here may be said to be ones of the {\it first-order}.

On the other hand, the finer evaluation of achievable rates, called the {\it second-order} achievable rates, have been investigated in several contexts. In variable-length source coding problem, Kontoyiannis \cite{Kon} has established the second-order source coding theorem. In channel coding problem, Strassen (see, Csisz\'{a}r and K\"{o}rner \cite{CK}), Hayashi \cite{Hayashi2}, and Polyanskiy, Poor and Verd\'{u} \cite{Polyanskiy2010}
have determined the second-order capacity rates.
%These analyses are based upon the asymptotic normality.
Hayashi \cite{Hayashi} has also shown the second-order achievability theorems for the intrinsic randomness problem as well as for the fixed-length source coding problem with {\it general} sources. %Hayashi \cite{Hayashi} also has revealed a relationship between these problems.
In addition, for i.i.d. sources he has demonstrated the calculation of the second-order optimal achievable rates by using the asymptotic normality in both problems.
%His result implies that in random number generation problem, we can deduce more precise results under the restriction for sources. 

In the present paper, we address the computation problem concerning the second-order formulas for resolvability, intrinsic randomness, and fixed-length source coding problems for {\it mixed} (non-i.i.d.) sources, where the resolvability problem was first and partly studied by Nomura and Matsushima \cite{Nomura_SITA2010}.
In the resolvability problem or the intrinsic randomness problem the degree of approximation is measured in terms of variational distance. 

As we have mentioned in the above, the analysis based upon the asymptotic normality is effective in deriving the second-order achievable rates. However, it had earlier been applied only to the class which has a simple probabilistic structure, such as  i.i.d. sources, Markovian sources or stationary discrete memoryless channels but {\it not} to that of {\it mixed} sources.

In the present paper, we first establish the resolvability formula for general sources, and then specifically compute the second-order optimal achievable rates for {\it mixed} sources, which is a wider but still tractable class of sources than the previous ones in random number generation. Recall that mixed sources are typical cases of {\it nonergodic} sources.
Nonetheless, we show that we can use still the {\it two-peak} asymptotic normality to compute the second-order achievable rates % in resolvability problem and source coding problem 
for mixed sources.

Related works include, e.g., Polyanskiy, Poor and Verd\'{u} \cite{Polyanskiy2011} that has developed the second-order capacity of the Gilbert-Elliott channel (GEC). % from the viewpoint of a two-peak asymptotic normality.
 The GEC is a simple model of channels in which the crossover probability of a {\it binary} symmetric channel obeys a {\it binary} symmetric Markov chain transition.
In particular, they have analyzed the {\it nonergodic} case in the GEC using the two-peak asymptotic normality from the viewpoint of a {\it mixture} of two memoryless channels.
On the other hand, in this paper we consider a mixed source consisting of two i.i.d sources with {\it countably infinite} alphabet. Our analysis is on the basis of the {\it information spectrum} methods, and hence valid with countably infinite alphabet. In addition, it is shown that our results can be easily extended to more general cases.
Actually, we generalize our results to mixed sources consisting of {\it countably infinite} i.i.d. sources with {\it countably infinite} alphabet, and furthermore to mixed sources consisting of {\it countably infinite} Markovian sources but with {\it finite} alphabet.
The resolvability formula for mixed sources with {\it general mixture} not necessarily countably infinite mixture is also established.

It should be emphasized here that we have recourse to a bulk of {\it information spectrum calculations} throughout in the paper. Although they apparently look tedious and even cumbersome, each step in the process of computations is actually simple and very basic, which features the information spectrum unifying approach.

This paper is organized as follows. In Section II, we review the previous results on the first-order asymptotics for {\it general} sources. In section III, we review and derive the second-order asymptotic formulas for the {\it general} sources, analogously to Section II. In Section IV, we define the mixed source and state the lemmas which play the key role in the subsequent analysis.
In Sections V, VI and VII, with {\it mixed} sources we establish the second-order achievability by invoking the {\it two-peak} asymptotic normality, for the resolvability problem, the intrinsic randomness problem and the fixed-length source coding problem, respectively.
 In Section VIII, we point out that all the results established in Sections V--VII are still valid if we consider more general mixed sources.
Finally, we conclude our results in Section IX.
%
%
%  Section 2
%
%
%
\section{First-Order Asymptotics}
In this section we review the previous results on the first-order asymptotics of random number generation and fixed-length source coding.

To this end, we first give the necessary notations and definitions.
In the sequel, let ${\bf Y} =  \{ Y^n \}_{n=1}^\infty$ be a general source with values in countable sets ${\cal Y}^n$.
Let ${\cal Z}$ be a countable set and let $Z$, $\overline{Z}$ be random variables with values in ${\cal Z}$. Denote by $d(Z, \overline{Z})$ the variational distance
$$
d(Z,{\overline{Z}}) \equiv \sum_{z \in {{\cal Z}}} \left| P_{Z}(z) - P_{\overline{Z}}(z) \right|,
$$
where $P_{X}(\cdot)$ denotes the probability distribution of random variable $X$. Moreover, set $\ {\cal U}_{M} \equiv \{1,2, \cdots, M \} $ and let $U_M$ denote the random variable uniformly distributed on ${\cal U}_M$.
%
% 2-A
%
\subsection{First-Order Resolvability}
\begin{definition}
Rate $R$ is said to be $\delta$-achievable if there exists a mapping $\phi_n : {\cal U}_{M_n} \to {\cal Y}^n$ such that 
$$
\limsup_{n \rightarrow \infty} \frac{1}{n} \log M_n \leq R \mbox{ and } \limsup_{n \rightarrow \infty} d(Y^n, \phi_n(U_{M_n})) \leq \delta.
$$
\end{definition}
\begin{definition} [$\delta$-resolvability]
\begin{eqnarray*} 
S_{r}(\delta|{\bf Y}) = \inf \left\{ R \left|R \mbox{ is $\delta$-achievable} \right. \right\}.
\end{eqnarray*}
\end{definition}

Then, we have
\begin{theorem}[Steinberg and Verd\'{u} \cite{Steinberg}] \label{theo:1-1}
\begin{equation*} \label{1st_resolvability}
S_{r}(\delta|{\bf Y}) = \inf \left\{ R \left| F(R) \leq \frac{\delta}{2} \right. \right\} \ (0 \leq \forall \delta < 2), 
\end{equation*}
where 
\begin{equation} \label{FR}
F(R) = \limsup_{n \to \infty} \Pr \left\{ \frac{1}{n} \log \frac{1}{P_{Y^n}(Y^n)} \geq R \right\}.
\end{equation}
\end{theorem}
The following Fig. \ref{first-order_resolvability} illustrates Theorem \ref{theo:1-1}.
\begin{figure}[h]
\begin{center}
\includegraphics[width=3.0in]{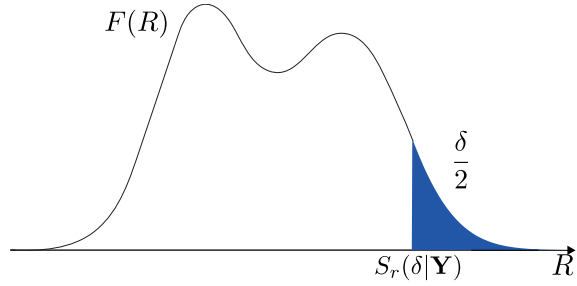}
% where an .eps filename suffix will be assumed under latex, 
% and a .pdf suffix will be assumed for pdflatex; or what has been declared
% via \DeclareGraphicsExtensions.
\centering \caption{First-Order Resolvability Rate}
\label{first-order_resolvability}
\end{center}
\end{figure}
%
% 2-B
%
\subsection{First-Order Intrinsic Randomness}
\begin{definition}
Rate $R$ is said to be $\delta$-achievable if there exists a mapping $\phi_n: {\cal Y}^n \to {\cal U}_{M_n}$ such that 
$$
\liminf_{n \rightarrow \infty} \frac{1}{n} \log M_n \geq R \mbox{ and } \limsup_{n \rightarrow \infty} d(U_{M_n},\phi_n(Y^n)) \leq \delta.
$$
\end{definition}
\begin{definition} [$\delta$-intrinsic randomness]
\begin{eqnarray*} 
S_{\iota}(\delta|{\bf Y}) = \sup \left\{ R \left|R \mbox{ is $\delta$-achievable} \right. \right\}.
\end{eqnarray*}
\end{definition}

Then, we have
\begin{theorem}[Han \cite{Han}] \label{theo:1-2}
\begin{equation*} \label{1st_ir}
S_{\iota}(\delta|{\bf Y}) = \sup \left\{ R \left| G(R) \leq \frac{\delta}{2} \right. \right\} \ (0 \leq \forall \delta < 2),
\end{equation*}
where 
\begin{equation*} \label{GR}
G(R) = \limsup_{n \to \infty} \Pr \left\{ \frac{1}{n} \log \frac{1}{P_{Y^n}(Y^n)} \leq R \right\}.
\end{equation*}
\end{theorem}
The following Fig. \ref{first-order ir} illustrates Theorem \ref{theo:1-2}.
\begin{figure}[h]
\begin{center}
\includegraphics[width=3.0in]{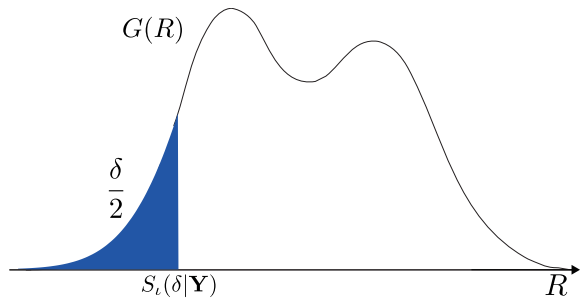}
% where an .eps filename suffix will be assumed under latex, 
% and a .pdf suffix will be assumed for pdflatex; or what has been declared
% via \DeclareGraphicsExtensions.
\centering \caption{First-Order Intrinsic Randomness Rate}
\label{first-order ir}
\end{center}
\end{figure}
%
% 2-C
%
\subsection{First-Order Fixed-length Source Coding}
Let $\varphi_n:{\cal Y}^n \to {\cal U}_{M_n}$, $\psi_n:{\cal U}_{M_n} \to {\cal Y}^n$ be an encoder and a decoder, respectively, for source ${\bf Y} = \{ Y^n \}_{n=1}^\infty$. 
The decoding error probability $\varepsilon_n$ is given by 
$
\varepsilon_n \equiv \Pr \left\{ Y^n \neq \psi_n(\varphi_n(Y^n)) \right\}.
$
Such a code is denoted by $(n, M_n, \varepsilon_n)$.
\begin{definition}
Rate $R$ is said to be $\varepsilon$-achievable if there exists a code $(n, M_n, \varepsilon_n)$ such that 
\[
\limsup_{n \rightarrow \infty} \varepsilon_n \leq \varepsilon \mbox{ and } \limsup_{n \rightarrow \infty}  \frac{1}{n} \log M_n \leq R.
\]
\end{definition}
\begin{definition}[$\varepsilon$-fixed-length source coding rate]
\begin{equation*}
L_{f}(\varepsilon|{\bf Y}) =  \inf \left\{ R | \mbox{$R$ is $\varepsilon$-achievable} \right\}.
\end{equation*}
\end{definition}

Then, we have
\begin{theorem}[Steinberg and Verd\'{u} \cite{Steinberg}] \label{theo:1-3}
\begin{equation*}
L_{f}(\varepsilon|{\bf Y}) = \inf \{ R | F(R) \leq \varepsilon \} \ (0 \leq \forall \varepsilon < 1),
\end{equation*}
where $F(R)$ is defined as in (\ref{FR}).
\end{theorem}
The following Fig. \ref{first-order sc} illustrates Theorem \ref{theo:1-3}.
\begin{figure}[hbp]
\begin{center}
\includegraphics[width=3.0in]{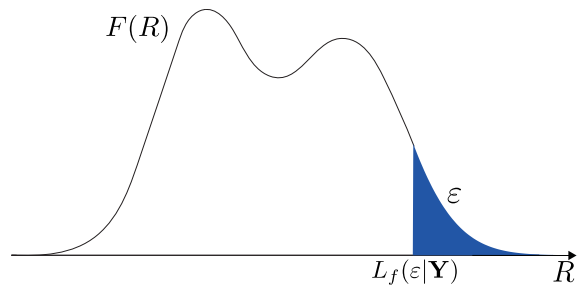}
% where an .eps filename suffix will be assumed under latex, 
% and a .pdf suffix will be assumed for pdflatex; or what has been declared
% via \DeclareGraphicsExtensions.
\centering \caption{First-Order Fixed-length Source Coding Rate}
\label{first-order sc}
\end{center}
\end{figure}

An immediate consequence of Theorem \ref{theo:1-1} and Theorem \ref{theo:1-3} is the following theorem, which reveals an operational equivalence between resolvability and fixed-length source coding from the viewpoint of random number generation, that is,
%\begin{theorem}[Han {[5, Remark 2.4.1]}]   \label{theo:1-4}
\begin{theorem}[{Han \cite[Remark 2.4.1]{Han}}]   \label{theo:1-4}
\begin{equation*}
L_{f}(\varepsilon|{\bf Y}) = S_{r}(2\varepsilon | {\bf Y}) \ (0 \leq \forall \varepsilon < 1).
\end{equation*}
\end{theorem}
As for the operational meaning of this equivalence, see Appendix \ref{app1}.
%
%
%
%  section 3
%
%
%
\section{Second-Order Asymptotics}
Having reviewed the results on the first-order asymptotics, we now focus on the second-order asymptotics, which will turn out to be in nice correspondence with the first-order asymptotics.
%
% 3-A
%
\subsection{Second-Order Resolvability}
\begin{definition} \label{def:3-1}
Rate $R$ is said to be $(a,\delta)$-achievable if there exists a mapping $\phi_n : {\cal U}_{M_n} \to {\cal Y}^n$ such that 
$$
\limsup_{n \rightarrow \infty} \frac{1}{\sqrt{n}} \log \frac{M_n}{e^{na}} \leq R \mbox{ and } \limsup_{n \rightarrow \infty} d(Y^n, \phi_n(U_{M_n})) \leq \delta.
$$
\end{definition}
\begin{definition} [$(a, \delta)$-resolvability] \label{def:3-2}
\begin{eqnarray*} 
S_{r}(a, \delta|{\bf Y}) = \inf \left\{ R \left|R \mbox{ is $(a, \delta)$-achievable} \right. \right\}.
\end{eqnarray*}
\end{definition}

Then, we obtain the following fundamental formula for the resolvability problem:
\begin{theorem} \label{theo:2-1}
\begin{equation*} 
S_{r}(a,\delta|{\bf Y}) = \inf \left\{ R \left| F_a(R) \leq \frac{\delta}{2} \right. \right\} \ (0 \leq \forall \delta < 2),
\end{equation*}
where 
\begin{equation} \label{FaR}
F_a(R) = \limsup_{n \to \infty} \Pr \left\{ \frac{1}{n} \log \frac{1}{P_{Y^n}(Y^n)} \geq a + \frac{R}{\sqrt{n}} \right\}.
\end{equation}
\end{theorem}
The following Fig. \ref{second-order resolvability} illustrates Theorem \ref{theo:2-1}.
\begin{figure}[!h]
\begin{center}
\includegraphics[width=3.4in]{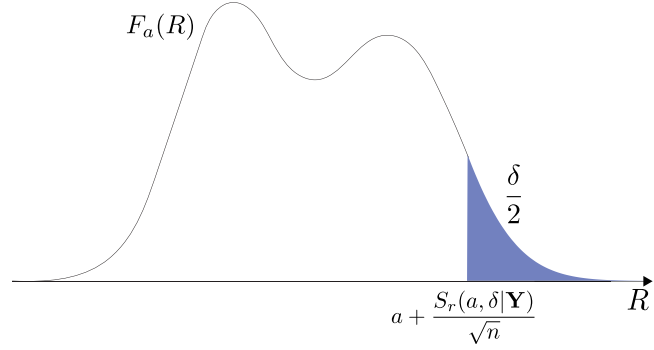}
% where an .eps filename suffix will be assumed under latex, 
% and a .pdf suffix will be assumed for pdflatex; or what has been declared
% via \DeclareGraphicsExtensions.
\centering \caption{Second-Order Resolvability Rate}
\label{second-order resolvability}
\end{center}
\end{figure}
\begin{remark}
This theorem can be driven as a consequence of Theorem \ref{theo:2-3} combined with Theorem \ref{theo:2-4} below. See also Remark \ref{remark3-2} in Subsection \ref{subsec:3} below.
\end{remark}
%
% 3-B
%
\subsection{Second-Order Intrinsic Randomness}
\begin{definition}
Rate $R$ is said to be $(a, \delta)$-achievable if there exists a mapping $\phi_n: {\cal Y}^n \to {\cal U}_{M_n}$ such that 
$$
\liminf_{n \rightarrow \infty} \frac{1}{\sqrt{n}} \log \frac{M_n}{e^{na}} \geq R \mbox{ and } \limsup_{n \rightarrow \infty} d(U_{M_n},\phi_n(Y^n)) \leq \delta.
$$
\end{definition}
\begin{definition} [$(a, \delta)$-intrinsic randomness]
\begin{eqnarray*} 
S_{\iota}(a, \delta|{\bf Y}) = \sup \left\{ R \left|R \mbox{ is $(a, \delta)$-achievable} \right. \right\}.
\end{eqnarray*}
\end{definition}

Then, we have
\begin{theorem}[Hayashi \cite{Hayashi}] \label{theo:2-2}
\begin{equation*} 
S_{\iota}(a,\delta|{\bf Y}) = \sup \left\{ R \left| G_a(R) \leq \frac{\delta}{2} \right. \right\} \ (0 \leq \forall \delta < 2),
\end{equation*}
where 
\begin{equation*} \label{GaR}
G_a(R) = \limsup_{n \to \infty} \Pr \left\{ \frac{1}{n} \log \frac{1}{P_{Y^n}(Y^n)} \leq a + \frac{R}{\sqrt{n}} \right\}.
\end{equation*}
\end{theorem}
The following Fig. \ref{second-order ir} illustrates Theorem \ref{theo:2-2}.
\begin{figure}[h]
\begin{center}
\includegraphics[width=3.2in]{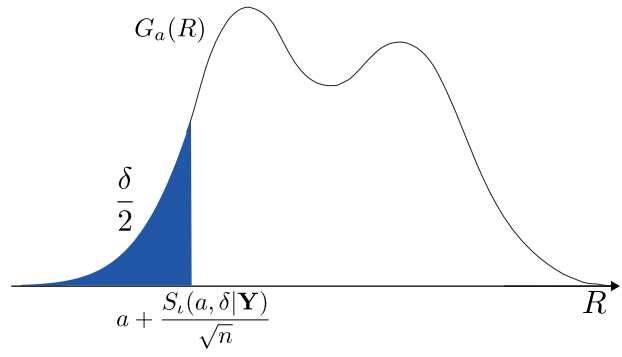}
% where an .eps filename suffix will be assumed under latex, 
% and a .pdf suffix will be assumed for pdflatex; or what has been declared
% via \DeclareGraphicsExtensions.
 \caption{Second-Order Intrinsic Randomness Rate}
\label{second-order ir}
\end{center}
\end{figure}
\clearpage
%3-C
%
\subsection{Second-Order Fixed-length Source Coding} \label{subsec:3}
\begin{definition}
Rate $R$ is said to be $(a, \varepsilon)$-achievable if there exists a code $(n, M_n, \varepsilon_n)$ such that 
$$
\limsup_{n \rightarrow \infty} \varepsilon_n \leq \varepsilon \mbox{ and } \limsup_{n \rightarrow \infty}  \frac{1}{\sqrt{n}} \log \frac{M_n}{e^{na}} \leq R.
$$
\end{definition}
\begin{definition}[$(a,\varepsilon)$-fixed-length source coding rate]
\begin{equation*}
L_{f}(a, \varepsilon|{\bf Y}) =  \inf \left\{ R | \mbox{$R$ is $(a, \varepsilon)$-achievable} \right\}.
\end{equation*}
\end{definition}

Then, we have
\begin{theorem}[Hayashi \cite{Hayashi}] \label{theo:2-3}
\begin{equation*}
L_{f}(a, \varepsilon|{\bf Y}) = \inf \{ R | F_a(R) \leq \varepsilon \} \ (0 \leq \forall \varepsilon < 1)
\end{equation*}
where $F_a(R)$ is defined as in (\ref{FaR}).
\end{theorem}
The following Fig. \ref{second-order sc} illustrates Theorem \ref{theo:2-3}.
\begin{figure}[h]
\begin{center}
\includegraphics[width=3.1in]{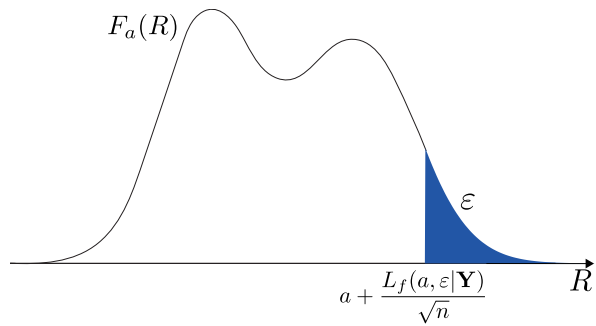}
% where an .eps filename suffix will be assumed under latex, 
% and a .pdf suffix will be assumed for pdflatex; or what has been declared
% via \DeclareGraphicsExtensions.
\centering \caption{Second-Order Fixed-length Source Coding Rate}
\label{second-order sc}
\end{center}
\end{figure}

On the other hand, we have the following equivalence theorem between resolvability and fixed-length source coding, which is a finer version of Theorem \ref{theo:1-4} for the case of the second-order asymptotics:
\vspace*{\baselineskip}
\begin{theorem}[equivalence theorem] \label{theo:2-4}
\begin{equation*}
L_{f}(a, \varepsilon|{\bf Y}) = S_{r}(a, 2\varepsilon | {\bf Y}) \ (0 \leq \forall \varepsilon < 1).
\end{equation*}
\end{theorem}
\begin{IEEEproof}
This theorem is to reveal a kind of {\it operational relationship} between resolvability and fixed-length source coding.
It suffices to show that if $R> L_f(a, \varepsilon|{\bf Y})$ then $R + \gamma > S_{r}(a, 2\varepsilon | {\bf Y})$ for any small $\gamma > 0$, and vice versa with $R$ and $R+\gamma$ swapped.
To this end, it is sufficient to literally follow the arguments described in Han \cite[p.163]{Han} with $\gamma$ replaced by $\frac{\gamma}{\sqrt{n}}$. 
To see its mechanism explicitly, we will give the details of the proof in Appendix \ref{app1}.
\end{IEEEproof}
\begin{remark} \label{remark3-2}
Theorem \ref{theo:2-1} can also directly, not via Theorem \ref{theo:2-4}, be proved by using Lemma \ref{lemma1} and Lemma \ref{lemma2} to be described in Section \ref{sec:5}.
Thus, the second-order resolvability formula for {\it general} sources can be reasonably established by using the argument similar to that for the first-order formula.
This is because the lemmas obtained by the {\it information spectrum} methods are very basic and fundamental.
This viewpoint has been pointed out also in Hayashi \cite{Hayashi}. 
Actually, in \cite{Hayashi} the second-order formula for intrinsic randomness (Theorem \ref{theo:2-2}) as well as for fixed-length source coding (Theorem \ref{theo:2-3}) has been proved using these lemmas (also cf. Lemmas \ref{lemma3} and \ref{lemma4} in Section VII), which had earlier been used already to establish the first-order formulas for these problems (cf.  Han \cite{Han}).
\end{remark}
%
%
% Section4
%
%
\section{Second-order asymptotics for mixed sources}
So far we have demonstrated the general formulas for typical first-order and second-order asymptotic problems (resolvability, intrinsic randomness and fixed-length source coding rate) of random number generation with any general source ${\bf Y} = \{ Y^n\}_{n=1}^\infty$.

However, computation of these general formulas is quite hard in general or even formidable. Therefore, in this section we consider to introduce a class of tractable sources ${\bf Y}$ for which the general formulas are computable but still of independent interest.
One of such source classes would be the case where ${\bf Y}$ is a {\it mixed} source of two i.i.d. sources ${\bf Y}_1 = \{ Y_1^n\}_{n=1}^\infty$ and ${\bf Y}_2 = \{ Y_2^n\}_{n=1}^\infty$.
The computation problem of the first-order asymptotics for such mixed sources has already been solved (e.g., see Han \cite{Han}), so in the sequel we now focus on the computation problem of the second-order asymptotics for mixed sources.
As a result, it will turn out that we can explicitly compute the asymptotic formulas by virtue of the {\it information spectrum methods}.

Let us begin with the formal definition of mixed sources.
Let ${\cal Y}= \{0,1,2, \cdots \}$ ({\it countably infinite}) be a source alphabet and ${\bf y} = y_1 y_2 \cdots y_n \in {\cal Y}^n $ denote a sequence emitted from the source of length $n$. Let $Y^n$ denote a random variable: a source sequence of length $n$. 

We consider a mixed source consists of two stationary memoryless sources ${\bf Y}_{i}
 = \{ Y^n_{i} \}_{n=1}^\infty$ with $i=1,2$.
Then, the mixed source ${\bf Y} = \{ Y^n \}_{n=1}^\infty$ is defined by
\begin{equation} \label{mixed}
P_{Y^n}({\bf y}) = w(1)P_{Y_{1}^n}({\bf y}) + w(2)P_{Y_{2}^n}({\bf y}),
\end{equation}
where $w(i)$ are  constants satisfying $w(1) + w(2) = 1$ and $w(i) > 0$ $(i=1,2)$.
Since two i.i.d. sources ${\bf Y}_i$ $(i =1,2)$ are completely specified by giving just the first component $Y_i$ $(i =1,2)$, we may write simply as ${\bf Y}_i = \{Y_i \}$ $( i=1,2)$ and define the variances:
\begin{definition}[variance] \label{var}
\begin{eqnarray*}
\sigma^2_i = E \left( \log \frac{1}{P_{Y_i}(Y_i)}- H(Y_i) \right)^2 \ (i=1,2) ,
\end{eqnarray*}
where we assume that these variances are finite, and define the entropy by
\[H(Y_i) = \sum_{{ y} \in {\cal Y}} P_{Y_i}({y}) \log \frac{1}{P_{Y_i}({ y})}.\]
\end{definition}
Since we consider the case where ${\bf Y}_i = \{Y_i \}$ $( i=1,2)$ is an i.i.d. source, the following asymptotic normality  holds for each component i.i.d. source:
%\begin{multline*}
%\lim_{n \rightarrow \infty}\Pr\left\{ \frac{-\log P_{Y_i^n}(Y_i^n) - nH(Y_i)}{\sqrt{n }{\sigma_i}} \leq U \right\} \\
% = \int_{-\infty}^{U} \frac{1}{\sqrt{2\pi}}\exp\left[ -\frac{z^2}{2} \right] dz,
\begin{equation} \label{normality}
\lim_{n \rightarrow \infty}\Pr\left\{ \frac{-\log P_{Y_i^n}(Y_i^n) - nH(Y_i)}{\sqrt{n }{\sigma_i}} \geq U \right\}  = \int^{+\infty}_{U} \frac{1}{\sqrt{2\pi}}\exp\left[ -\frac{z^2}{2} \right] dz,
\end{equation}
%\end{multline*}
where ${\sigma_i}^2$ denotes the variances defined in Definition \ref{var} $(i=1,2)$.

\vspace*{\baselineskip}
The following lemma plays the key role in dealing with {\it mixed} sources in the proof of Theorem \ref{theo1}, Theorem \ref{theo:IR} and Theorem \ref{theo2}. In other words, the crux of the arguments for mixed sources in the present paper is summarized by Lemma \ref{lemma}, which is completely irrelevant to the matters posed by Hayashi \cite{Hayashi}.
\begin{lemma}[Han \cite{Han}]\label{lemma}
Let $\{ z_n \}_{n=1}^\infty$ be any real-valued sequence.
Then for the mixed source ${\bf Y}$ it holds that, for $i = 1,2$,
\begin{enumerate}
\item %\begin{eqnarray*}
%\lefteqn{ \Pr\left\{ \frac{-\log P_{Y^n}(Y^n_{i}) }{\sqrt{n}}\geq z_n \right\} } \nonumber \\
%&\geq& \Pr\left\{ \frac{-\log P_{Y^n_{i}}(Y^n_{i}) }{\sqrt{n}} \geq z_n + \gamma_n \right\} - e^{-\sqrt{n}\gamma_n} ,
\[
 \Pr\left\{ \frac{-\log P_{Y^n}(Y^n_{i}) }{\sqrt{n}}\geq z_n \right\}  \geq  \Pr\left\{ \frac{-\log P_{Y^n_{i}}(Y^n_{i}) }{\sqrt{n}} \geq z_n + \gamma_n \right\} - e^{-\sqrt{n}\gamma_n},
 \]
 %\end{eqnarray*}
\item %\begin{eqnarray*}
%\lefteqn{ \Pr\left\{ \frac{-\log P_{Y^n}(Y^n_{i}) }{\sqrt{n}}\geq z_n \right\} } \nonumber \\
%&\leq& \Pr\left\{ \frac{-\log P_{Y^n_{i}}(Y^n_{i}) }{\sqrt{n}} \geq z_n - \gamma_n \right\} ,
\[
\Pr\left\{ \frac{-\log P_{Y^n}(Y^n_{i}) }{\sqrt{n}}\geq z_n \right\} \leq \Pr\left\{ \frac{-\log P_{Y^n_{i}}(Y^n_{i}) }{\sqrt{n}} \geq z_n - \gamma_n \right\} , 
%\hspace*{4.25zw}
\]
%\end{eqnarray*}
\end{enumerate}
where $\gamma_n > 0$ satisfies $\gamma_1 > \gamma_2 > \cdots > 0,$ $\gamma_n \to 0$, $\sqrt{n} \gamma_n \to \infty$.
\end{lemma}
\IEEEproof See Appendix \ref{app2}. \IEEEQED
%
%
% Section 5
%
%
\section{$(a,\delta)$-resolvability } \label{sec:5}
In this section we shall establish $S_r(a,\delta|{\bf Y})$ for mixed sources.
At first we introduce here two fundamental lemmas of Han \cite{Han}:
%The infimum of $d$-achievable rates has been determined by using two lemmas in Han \cite{Han}. 
Lemma \ref{lemma1} and Lemma \ref{lemma2} below.
Before describing the lemmas, we need to define two sets. 
Let ${\bf X} = \{ X^n\}_{n=1}^\infty$ and ${\bf Y}= \{ Y^n\}_{n=1}^\infty$ be arbitrary general sources with {\it countably infinite} alphabets, and given a sequence $\{ z_n\}_{n=1}^\infty$ define  $S_n(z_n)$ and $T_n(z_n)$:
\begin{eqnarray*}
S_n(z_n) = \left\{ {\bf x} \in {\cal X}^n \left| \frac{1}{\sqrt{n}}\log\frac{1}{P_{X^n}({\bf x})} \geq z_n \right. \right\},
\end{eqnarray*}
\begin{eqnarray*}
T_n(z_n) = \left\{ {\bf y} \in {\cal Y}^n \left| \frac{1}{\sqrt{n}}\log\frac{1}{P_{Y^n}({\bf y})} \leq z_n \right. \right\}.
\end{eqnarray*}
\begin{lemma} \label{lemma1}
Let ${\bf X} = \{ X^n\}_{n=1}^\infty$ and ${\bf Y}= \{ Y^n\}_{n=1}^\infty$ be arbitrary general sources, where $X^n$ and $Y^n$ are random variables taking values in ${\cal X}^n$ and ${\cal Y}^n$, respectively. Then, for an arbitrary sequence $\{z_n\}_{n=1}^\infty$ and $\gamma > 0$, there exists a mapping $\phi_n:{\cal X}^n \to {\cal Y}^n$ such that
\begin{eqnarray*}
%\lefteqn{d(\phi_n(X^n),Y^n)} \\
%& \leq & 2 \max\left( \Pr\{X^n \notin S_n(z_n\!+\!\gamma) \},\Pr\{Y^n \notin T_n(z_n) \}  \right) \\
%&      & + 2 e^{-\sqrt{n}\gamma}. 
d(\phi_n(X^n),Y^n) \leq 2 \max\left( \Pr\{X^n \notin S_n(z_n\!+\!\gamma) \},\Pr\{Y^n \notin T_n(z_n) \}  \right) + 2 e^{-\sqrt{n}\gamma}. 
\end{eqnarray*} 
\end{lemma} 
\begin{lemma} \label{lemma2}
Let ${\bf X} = \{ X^n\}_{n=1}^\infty$ and ${\bf Y}= \{ Y^n\}_{n=1}^\infty$ be arbitrary general sources, where $X^n$ and $Y^n$ are random variables taking values in ${\cal X}^n$ and ${\cal Y}^n$, respectively. Then, for an arbitrary sequence $\{z_n\}_{n=1}^\infty$, $\gamma > 0$ and any mapping $\phi_n:{\cal X}^n \to {\cal Y}^n$ it holds that
\begin{eqnarray} \label{eq:5-1}
%\lefteqn{d(\phi_n(X^n),Y^n)} \\
%& \geq & 2 \Pr\{Y^n \notin T_n(z_n\!+\!\gamma) \} - 2 \Pr\{X^n \in S_n(z_n) \} \\
%&      & - 2 e^{-\sqrt{n}\gamma},
d(\phi_n(X^n),Y^n) \geq 2 \Pr\{Y^n \notin T_n(z_n\!+\!\gamma) \} - 2 \Pr\{X^n \in S_n(z_n) \} - 2 e^{-\sqrt{n}\gamma}.
\end{eqnarray} 
\end{lemma}
\begin{remark}
Also, (\ref{eq:5-1}) can be written as
\begin{equation} \label{eq:5-2}
d(\phi_n(X^n),Y^n) \geq 2 \Pr\{X^n \notin S_n(z_n) \} - 2 \Pr\{Y^n \in T_n(z_n + \gamma) \} -  2 e^{-\sqrt{n}\gamma}.
\end{equation}
\end{remark}

The above lemmas are useful for the random number generation problem to approximate a probability distribution ${\bf Y}= \{ Y^n\}_{n=1}^\infty$ by using an another probability distribution ${\bf X} = \{ X^n\}_{n=1}^\infty$. Clearly, this includes the resolvability problem as a special case, which is the case of $X^n = U_{M_n}$.
Therefore, in this case the condition in the above lemmas leads to
\begin{eqnarray} \label{reduced}
\Pr\{ X^n \notin S_n(z_n) \} =  \left\{
   \begin{array}{ll}
    0 & z_n \leq \frac{1}{\sqrt{n}} \log M_n \\
    1 & z_n > \frac{1}{\sqrt{n}} \log M_n.
   \end{array} \right.
\end{eqnarray}
%holds in the resolvability problem and we use this expression.
Notice that the above lemmas are valid for general sources ${\bf X}$ and ${\bf Y}$.

\begin{figure}[b]
\begin{center}
\includegraphics[width=3.3in]{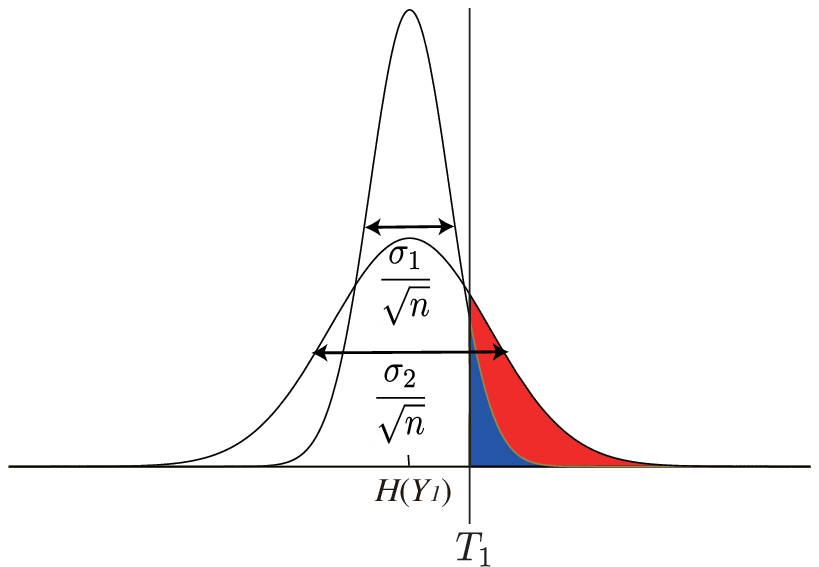}
% where an .eps filename suffix will be assumed under latex, 
% and a .pdf suffix will be assumed for pdflatex; or what has been declared
% via \DeclareGraphicsExtensions.
\centering \caption{Case I}
\label{mixture1}
\end{center}
\end{figure}

\begin{figure}
\begin{center}
\includegraphics[width=3.3in]{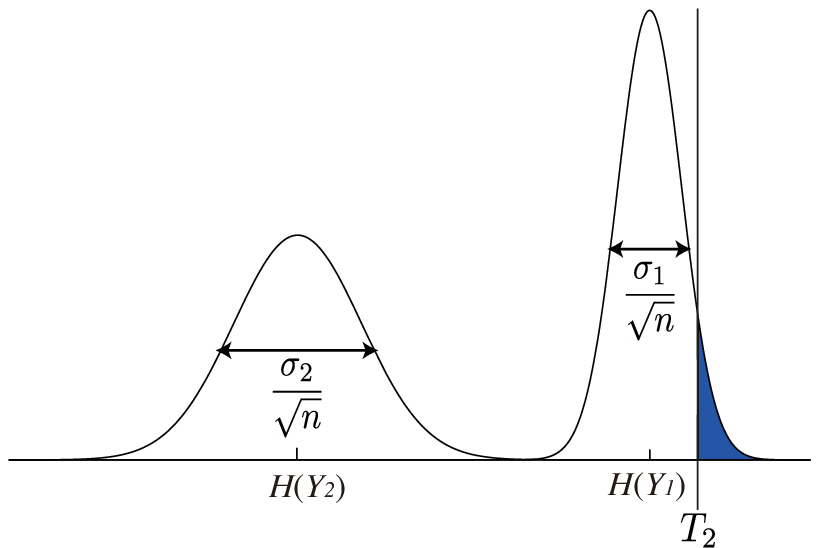}
% where an .eps filename suffix will be assumed under latex, 
% and a .pdf suffix will be assumed for pdflatex; or what has been declared
% via \DeclareGraphicsExtensions.
\caption{Case II}
\label{mixture2}
\end{center}
\end{figure}

\begin{figure}
\begin{center}
\includegraphics[width=3.3in]{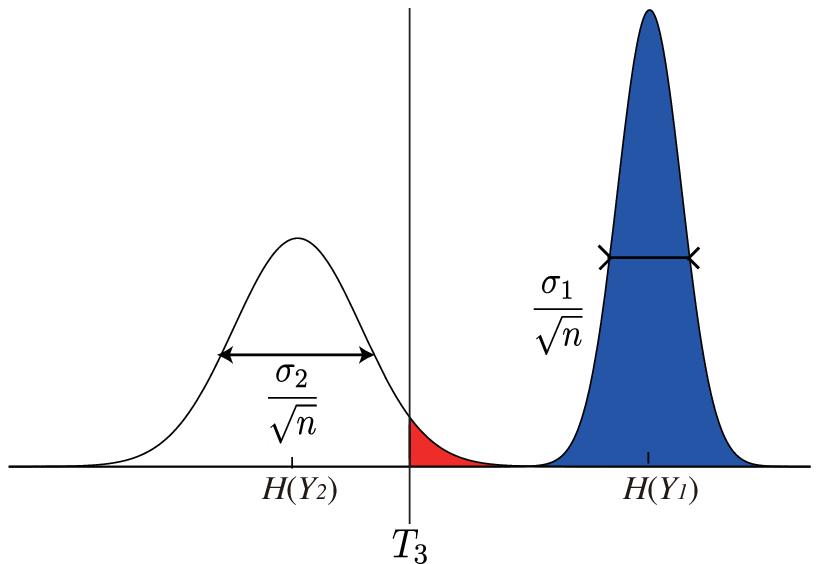}
% where an .eps filename suffix will be assumed under latex, 
% and a .pdf suffix will be assumed for pdflatex; or what has been declared
% via \DeclareGraphicsExtensions.
\caption{Case III}
\label{mixture3}
\end{center}
\end{figure}

In the sequel, we consider the case that $0 \leq \delta <2$ and $w(1) \neq \frac{\delta}{2}$ hold (cf. Remark \ref{remark1} below for the case of $w(1) = \frac{\delta}{2}$ ).
Then, given $0 \leq \delta < 2$ we classify the problem into three cases. Here, without loss of generality, we assume that $H(Y_1) \geq H(Y_2)$ holds:

\begin{description}
\item[I] $H(Y_1) = H(Y_2)$ holds.
\item[II] $H(Y_1) > H(Y_2)$ and $w(1) > \frac{\delta}{2}$ hold.
\item[III]	$H(Y_1) > H(Y_2)$ and $w(1) < \frac{\delta}{2}$ hold.
\end{description}
In Case I, we shall establish $S_{r}(H(Y_1),\delta|{\bf Y})$ (obviously, this is equal to $S_{r}(H(Y_2),\delta|{\bf Y})$). In Case II and Case III we shall show $S_{r}(H(Y_1),\delta|{\bf Y})$ and  $S_{r}(H(Y_2),\delta|{\bf Y})$, respectively. Now we have one of the main results:
\vspace*{\baselineskip}
\begin{theorem}\label{theo1}
Given $0 \leq \delta <2$, the following holds.
%\begin{description}

{\it Case I:}
\begin{eqnarray} \label{eq:main5-1}
S_{r}(H(Y_1),\delta|{\bf Y}) = T_1,
\end{eqnarray}

where $T_1$ is specified by
\begin{eqnarray} \label{T1}
\frac{\delta}{2} =  \sum_{i=1}^2 w(i)\int^{\infty}_{\frac{T_1}{\sigma_i}} \frac{1}{\sqrt{2\pi}}\exp\left[ -\frac{z^2}{2} \right] dz.
\end{eqnarray}

{\it Case II:}
\begin{eqnarray} \label{eq:main5-2}
S_{r}(H(Y_1),\delta|{\bf Y}) = T_2,
\end{eqnarray}

where $T_2$ is specified by
\begin{eqnarray} \label{T2}
\frac{\delta}{2} =  w(1) \int^{\infty}_{\frac{T_2}{\sigma_1}} \frac{1}{\sqrt{2\pi}}\exp\left[ -\frac{z^2}{2} \right] dz.
\end{eqnarray}

{\it Case III:}
\begin{eqnarray} 
S_{r}(H(Y_2),\delta|{\bf Y}) = T_3,
\end{eqnarray}

where $T_3$ is specified by
\begin{eqnarray} 
\frac{\delta}{2} =  w(1) + w(2) \int^{\infty}_{\frac{T_3}{\sigma_2}} \frac{1}{\sqrt{2\pi}}\exp\left[ -\frac{z^2}{2} \right] dz.
\end{eqnarray}
\end{theorem}
Illustrative figures in each case of this theorem are depicted in Fig. \ref{mixture1}--Fig. \ref{mixture3}, where the weighted probability of the shaded area is equal to $\frac{\delta}{2}$.
\newpage
\begin{remark} \label{remark1}
It is easy to check that $T_2= -\infty$, $T_3 = +\infty$ for $w(1) = \frac{\delta}{2}$.
Also, it will turn out from the way of proving the above theorem that the second-order asymptotics gets trivial if $a \neq H(Y_1)$ and $a \neq H(Y_2)$, because this case necessarily implies that $\delta=0$ or $\delta = 2w(1)$ or $\delta=2$, depending on the value of $a$; then, accordingly, we can formally set as $S_{r}(a,\delta|{\bf Y}) = - \infty$.
% for $a$ such that $a \neq H(Y_1)$ and $a \neq H(Y_2)$ with jointly achievable $\delta$'s.
\end{remark}
\begin{remark} \label{remark5-4}
Theorem \ref{theo1} can be restated more intuitively but equivalently as follows. % (cf. Polyanskiy \cite{Polyanskiy2011}).
Let
\[
R_n \equiv \frac{1}{n}\log M_n
\]
denote the {\it size rate} of resolvability (cf. Definition \ref{def:3-1} and \ref{def:3-2}), and consider the following asymptotic equation for $R_n$:
\begin{align} \label{solution1}
w(1)\Phi \left( \frac{\sqrt{n}(R_n - H(Y_1))}{\sigma_1} \right) + w(2)\Phi \left( \frac{\sqrt{n}(R_n - H(Y_2))}{\sigma_2} \right) = \frac{\delta}{2},
\end{align}
where $\Phi(\cdot)$ is the Gaussian cumulative distribution function defined by
\[
\Phi(x) = \frac{1}{\sqrt{2\pi}} \int_{x}^{+\infty} e^{-\frac{x^2}{2}} dx.
\]
Denote the solution of this equation by
\[
R_n^{\ast} = \frac{1}{n}\log M_n^{\ast}
\]
and  set as 
\[
R_n^{\ast} = a + \frac{b}{\sqrt{n}} + o\left(\frac{1}{\sqrt{n}}\right) \ (a, b \mbox{ are constants}) ,
\]
which, substituted into (\ref{solution1}), yields
\begin{align} \label{solution2}
w(1)\Phi \left( \frac{\sqrt{n}(a - H(Y_1)) + b + o(1)}{\sigma_1} \right) + w(2)\Phi \left( \frac{\sqrt{n}(a - H(Y_2)) + b +o(1)}{\sigma_2} \right) = \frac{\delta}{2}.
\end{align}
Then, it is not difficult to verify by letting $n \to \infty$ that, given $a$ and $\delta$, the corresponding solution $b=b^{\ast}(a,\delta)$ of this equation coincides with $T_1$, $T_2$ and $T_3$, respectively, according to Cases I, II, III (cf. also (\ref{eq:8-1-1}) in the proof of Theorem \ref{theo:8-1}).
Notice here that the equation (\ref{solution1}) subsumes Remark \ref{remark1} too. Thus, it is concluded that $b^{\ast}(a,\delta)$ is nothing but the second-order resolvability $S_r(a,\delta|{\bf Y})$, and hence Theorem \ref{theo1} is equivalent to the equation (\ref{solution1}).

Summarizing up, we can write the optimal size $M^{\ast}_n$ as 
\[
\log M^{\ast}_n = na + \sqrt{n}b^{\ast}(a,\delta) + o({\sqrt{n}}),
\]
with $a=H(Y_1)$ or $a = H(Y_2)$ depending on the value of $\delta$, that is, if $H(Y_1) \neq H(Y_2)$ then
\begin{eqnarray} \label{eq:pol2} \begin{array}{ll}
a = H(Y_1) \mbox{ and } b^{\ast}(a,\delta) = \sigma_1 \Phi^{-1}\left( \frac{\delta}{2 w(1)}\right) & \mbox{ if } w(1) > \frac{\delta}{2} ;  \\
a = H(Y_2) \mbox{ and } b^{\ast}(a,\delta) = \sigma_2 \Phi^{-1}\left( \frac{\delta - 2w(1)}{2 w(2)}\right) & \mbox{ if } w(1) < \frac{\delta}{2},
\end{array}
\end{eqnarray}
which enables us to evaluate how large size of $M^{\ast}_n$ is needed as a function of block length $n$ and variational distance $\delta$. Notice here that in this case $b^{\ast}(a,\delta)$ can be written as the simple inverse function $\Phi^{-1}$ of the Gaussian distribution function, and also that $b^{\ast}(a,\delta)$ can be negative, for example, $b^{\ast}(a,\delta) < 0 $ if $\delta > 1 + w(1)$, so that in this case the first-order $\delta$-resolvability $S_r(\delta|{\bf Y})$ is $H(Y_2)$ but the optimal achievable rate $R^{\ast}_n$ approaches it {\it from below}. 
In other words, it is possible to make necessary rates to be below the $\delta$-resolvability at finite block length $n$.
The nonergodic channel counterpart of the equations (\ref{solution1}) and (\ref{eq:pol2}) has been provided by Polyanskiy, Poor and Verd\'{u} \cite{Polyanskiy2011}, who have observed for the Gilbert Elliott channel the same kind of non-asymptotic phenomenon as here.
On the other hand, in the case where $H(Y_1) = H(Y_2)$ holds, $b^{\ast}(a,\delta)$ can be written as the inverse function of a mixed Gaussian distribution function (see, Case I). Notice here that Case I is missing in \cite{Polyanskiy2011}.
\IEEEQED
\end{remark}
%\begin{remark}
%Let us consider the case in Theorem \ref{theo1}, where ${\bf Y}$ is an i.i.d. source. This case can be regarded as the one with $w(1)=1$ and ${\bf Y} = \{Y_1\}$.
%Then, the theorem is rewritten as 
%\begin{eqnarray*}
%S_{r}(H(Y_1),\delta|{\bf Y}) = T,
%\end{eqnarray*}
%where $T$ is specified by
%\begin{eqnarray*}
%\frac{\delta}{2} =  \int^{\infty}_{\frac{T}{\sigma_1}} \frac{1}{\sqrt{2\pi}}\exp\left[ -\frac{z^2}{2} \right] dz.
%\end{eqnarray*}
%\end{remark}
%
% lemma
%
%To proof the above theorem we use the following lemma.
%
% 
%
%  proof of theorem
%
%
%
\begin{IEEEproof}[Proof of Theorem \ref{theo1}]
See Appendix \ref{app4}.
\end{IEEEproof}
%
%
% section 6
%
%
\section{$(a,\delta)$-intrinsic randomness}
Let us now turn to the computation problem of the $(a, \delta)$-intrinsic randomness formula for mixed sources. To do so, without loss of generality, we consider the following three cases:
\begin{description}
\item[I] $H(Y_1) = H(Y_2)$ holds.
\item[II] $H(Y_1) > H(Y_2)$ and $w(2) > \frac{\delta}{2}$ hold.
\item[III]	$H(Y_1) > H(Y_2)$ and $w(2) < \frac{\delta}{2}$ hold.
\end{description}
%In Case I, we shall establish $S_{r}(H(Y_1),\delta|{\bf Y})$ (Obviously this is equal to $S_{r}(H(Y_2),\delta|{\bf Y})$). In Case II and Case III we shall show $S_{r}(H(Y_1),\delta|{\bf Y})$ and  $S_{r}(H(Y_2),\delta|{\bf Y})$ respectively. Now we have the main results:
\vspace*{\baselineskip}

Then, we have
\begin{theorem}\label{theo:IR}
Given $0 \leq \forall \delta <2$, the following holds.
%\begin{description}

{\it Case I:}
\begin{eqnarray*} 
S_{\iota}(H(Y_2),\delta|{\bf Y}) = T_4,
\end{eqnarray*}

where $T_4$ is specified by
\begin{eqnarray*} 
\frac{\delta}{2} =  \sum_{i=1}^2 w(i)\int_{-\infty}^{\frac{T_4}{\sigma_i}} \frac{1}{\sqrt{2\pi}}\exp\left[ -\frac{z^2}{2} \right] dz.
\end{eqnarray*}

{\it Case II:}
\begin{eqnarray*}
S_{\iota}(H(Y_2),\delta|{\bf Y}) = T_5,
\end{eqnarray*}

where $T_5$ is specified by
\begin{eqnarray*} 
\frac{\delta}{2} =  w(2) \int_{-\infty}^{\frac{T_5}{\sigma_2}} \frac{1}{\sqrt{2\pi}}\exp\left[ -\frac{z^2}{2} \right] dz.
\end{eqnarray*}

{\it Case III:}
\begin{eqnarray*} 
S_{\iota}(H(Y_1),\delta|{\bf Y}) = T_6,
\end{eqnarray*}

where $T_6$ is specified by
\begin{eqnarray*} 
\frac{\delta}{2} =  w(2) + w(1) \int_{-\infty}^{\frac{T_6}{\sigma_1}} \frac{1}{\sqrt{2\pi}}\exp\left[ -\frac{z^2}{2} \right] dz.
\end{eqnarray*}
\end{theorem}
%
%  proof of theorem
%
\begin{IEEEproof} It suffices to proceed in parallel with the arguments as made in the proof of Theorem \ref{theo1}, while taking account of the {\it duality} between resolvability and intrinsic randomness. Note that (\ref{eq:5-2}) is used instead of (\ref{eq:5-1}) in the proof of Converse Part.
\end{IEEEproof}
\begin{remark}
It is easy to see that $T_5 = + \infty$ and $T_6 = -\infty$ for $w(2) = \frac{\delta}{2}$. The latter part of Remark \ref{remark1} similarly applies here too with $S_{\iota}(a,\delta|{\bf Y}) = + \infty$ instead of $S_{r}(a,\delta|{\bf Y}) = -\infty$.
Also, like in Remark \ref{remark5-4}, we can ascertain that Theorem \ref{theo:IR} is equivalent to the asymptotic equation:
\begin{align*}
w(1)\Psi \left( \frac{\sqrt{n}(R_n - H(Y_1))}{\sigma_1} \right) + w(2)\Psi \left( \frac{\sqrt{n}(R_n - H(Y_2))}{\sigma_2} \right) = \frac{\delta}{2},
\end{align*}
where $\Psi(x) = 1- \Phi(x)$ and $T_4$, $T_5$, $T_6$ instead of $T_1$, $T_2$, $T_3$, respectively.
\end{remark}
%
%
% Section 7
%
%
\section{$(a,\varepsilon)$-fixed-length source coding}
%
%
%\subsection{Relationship with Fixed-Length Source Coding}
Let us now consider to compute the formula for $L_{f}(a,\varepsilon|{\bf Y})$ for mixed sources. 
To do so, without loss of generality, we consider the following three cases:
%We shall determine $L_{F}(a, D|{\bf Y})$. Then, from Theorem \ref{theorem5} it coincides with $L_{dd}(a, D|{\bf Y})$.
%Again, we divide the problem into three cases.
\begin{description}
\item[I] $H(Y_1) = H(Y_2)$ holds.
\item[II] $H(Y_1) > H(Y_2)$ and $w(1) > \varepsilon$ hold.
\item[III]	$H(Y_1) > H(Y_2)$ and $w(1) < \varepsilon $ hold.
\end{description}
%In Case I, we shall show the formula for $L_{f}(H(Y_1),\varepsilon|{\bf Y})$ (Obviously this is equal to $L_{f}(H(Y_2),\varepsilon|{\bf Y})$). In Case II and Case III we shall show $L_{f}(H(Y_1),\varepsilon|{\bf Y})$ and  $L_{f}(H(Y_2),\varepsilon|{\bf Y})$ respectively. 
Then, we have the following main result:
\vspace*{\baselineskip}
\begin{theorem}\label{theo2}
Given $0 \leq \varepsilon < 1$, the following holds.

{\it Case I:}
\begin{eqnarray} \label{eq:main6-1}
L_{f}(H(Y_1),\varepsilon|{\bf Y}) = T_7,
\end{eqnarray}

where $T_7$ is specified by
\begin{eqnarray} \label{T4}
\varepsilon = \sum_{i=1}^2 w(i)\int^{\infty}_{\frac{T_7}{\sigma_i}} \frac{1}{\sqrt{2\pi}}\exp\left[ -\frac{z^2}{2} \right] dz.
\end{eqnarray}

{\it Case II:}
\begin{eqnarray} \label{eq:main6-2}
L_{f}(H(Y_1),\varepsilon|{\bf Y}) = T_8,
\end{eqnarray}

where $T_8$ is specified by
\begin{eqnarray*} \label{T8}
\varepsilon =  w(1) \int^{\infty}_{\frac{T_8}{\sigma_1}} \frac{1}{\sqrt{2\pi}}\exp\left[ -\frac{z^2}{2} \right] dz.
\end{eqnarray*}

{\it Case III:}
\begin{eqnarray*}  
L_{f}(H(Y_2),\varepsilon|{\bf Y}) = T_9,
\end{eqnarray*}

where $T_9$ is specified by
\begin{eqnarray*} 
\varepsilon =  w(1)\!+\!w(2) \int^{\infty}_{\frac{T_9}{\sigma_2}} \frac{1}{\sqrt{2\pi}}\exp\left[-\frac{z^2}{2} \right] dz.
\end{eqnarray*}
\end{theorem}
\begin{remark}
It is easy to check that $T_8= -\infty$, $T_9 = +\infty$ for $w(1) = \varepsilon$. The latter part of Remark \ref{remark1} similarly applies here too with $ \varepsilon$ instead of $\frac{\delta}{2}$ and $L_f(a,\varepsilon|{\bf Y}) = -\infty$  instead of $S_{r}(a,\delta|{\bf Y}) = -\infty$. Remark \ref{remark5-4} applies also here with $\varepsilon$ instead of $\frac{\delta}{2}$, and $T_7$, $T_8$, $T_9$ instead of $T_1$, $T_2$, $T_3$, respectively. In particular, it turns out that Theorem \ref{theo2} is equivalent to the equation (\ref{solution1}).
\end{remark}
\begin{IEEEproof}[Proof of Theorem \ref{theo2}] Although the proof is immediate from Theorem \ref{theo:2-4} and Theorem \ref{theo1} with $\varepsilon = \frac{\delta}{2}$, we give in Appendix \ref{app3} another information spectrum approach that is of independent interest, where Lemma \ref{lemma3} and Lemma \ref{lemma4} (due to Han \cite{Han}) as described below are invoked, which Hayashi \cite{Hayashi} has fully used to obtain the {\it general} formula for the second-order optimal fixed-length source coding rate (see, Theorem \ref{theo:2-3}) but not for {\it mixed} sources.
\end{IEEEproof}
%\vspace*{\baselineskip}
\begin{lemma}\label{lemma3}
Let $M_n$ be an arbitrary given positive integer. Then, for all $n = 1,2,\cdots,$ there exists an $(n,M_n,\varepsilon_n)$ code such that
\begin{eqnarray*}
\varepsilon_n \leq \Pr \left\{ \frac{1}{\sqrt{n}} \log\frac{1}{P_{Y^n}(Y^n)} \geq \frac{1}{\sqrt{n}} \log M_n \right\}.
\end{eqnarray*}
\end{lemma}
\begin{lemma}\label{lemma4}
For all $n=1,2,\cdots,$ any $(n, M_n, \varepsilon_n)$ code satisfies
\begin{equation*}
\varepsilon_n \geq  \Pr \left\{ \frac{1}{\sqrt{n}} \log\frac{1}{P_{Y^n}(Y^n)} \geq \frac{1}{\sqrt{n}} \log M_n\!+\!\gamma \right\}\!-\!e^{-\sqrt{n}\gamma},
\end{equation*}
where $\gamma > 0$ is an arbitrary constant.
\end{lemma}
%
%
%
%  Section 8
%
%
\section{Second-Order Resolvability for mixed Sources: General cases}
In the previous sections, we have established the second-order achievable rates for a mixture of two i.i.d. sources, which is regarded as being the simplest model of mixed sources.
%So, in this section, we extend our results to the 
%
In this section, we shall extend our results to more general mixed sources. 
Although we consider only the resolvability problem in this section,  similar extensions can be done immediately for the intrinsic randomness problem and the fixed-length coding problem.
\subsection{$(a,\delta)$-Resolvability for a Mixture of Countably Infinite i.i.d. Sources}
Let us now consider the mixed source consisting of {\it countably infinite} stationary memoryless sources ${\bf Y}_i = \{Y_i^n \}_{n=1}^{\infty}$, where $i \in \mathbb{Z} = \{0, \pm1, \pm2, \cdots \}$.
The mixed source that we consider in this subsection is defined by
\[
P_{Y^n}({\bf y}) = \sum_{i=-\infty}^\infty w(i) P_{Y^n_i}({\bf y}),
\]
where $w(i)$ are constants such that $\sum_{i = -\infty}^\infty w(i) =1$ and $w(i) \geq 0$ for all $i \in \mathbb{Z}$.
The variance $\sigma_i^2 \ (i \in \mathbb{Z})$ of the stationary memoryless source ${\bf Y}_i$ is defined in a similar way
to Definition \ref{var} and we assume that all these variances are finite.
Finally, without loss of generality we assume that
\[
\cdots \leq H(Y_{-2}) \leq H(Y_{-1}) \leq H(Y_0) \leq H(Y_{1}) \leq H(Y_2) \leq \cdots. 
\]
With this definition of mixed sources, we now have the following second-order resolvability theorem, which is a substantial generalization of Theorem \ref{theo1}.
%
%
%   theorem 8.1
%
%
\begin{theorem} \label{theo:8-1}
For a mixture ${\bf Y}$ of {\it countably infinite} i.i.d. sources with {\it countably infinite} alphabet, the optimal size rate 
$
R_n^{\ast} =\frac{1}{n} \log M^{\ast}_n
$
is given as the solution for $R_n$ of the asymptotic equation:
\begin{align} \label{solution3}
\sum_{i=-\infty}^{\infty} w(i)\Phi \left( \frac{\sqrt{n}(R_n - H(Y_i))}{\sigma_i} \right)= \frac{\delta}{2};
\end{align}
and, furthermore, the resolvability $S_r(a,\delta|{\bf Y})$
is given as the solution $b = b^{\ast}(a, \delta)$ of the asymptotic equation
\begin{align} \label{solution4}
\sum_{i=-\infty}^{\infty} w(i)\Phi \left( \frac{\sqrt{n}(a - H(Y_i))+ b + o(1)}{\sigma_i} \right)= \frac{\delta}{2}.
\end{align}
\end{theorem}
\begin{IEEEproof}
The derivation of equation (\ref{solution3}) and (\ref{solution4}) is based on the {\it multi-peak} asymptotic normality instead of the two-peak asymptotic normality.
Although it is easy to verify that Theorem \ref{theo:8-1} follows in a manner similar to that in Remark \ref{remark5-4}, for the sake of reader's convenience we demonstrate here also the formula of the form as in Theorem \ref{theo1}.
Define the sets of indices as
\[
{\cal L}_0(a) \equiv \{k \in \mathbb{Z}|{H}(Y_k) = a, \ w(k) >0 \}.
\]
\[
{\cal L}_1(a) \equiv \{i \in \mathbb{Z}|{H}(Y_i) > a, \ w(i) >0\},
\]
We first consider the case where ${\cal L}_0(a) = \emptyset$ (the empty set).
This case necessarily implies
\[
\sum_{i \in {\cal L}_1(a)} w(i) = \frac{\delta}{2},
\]
which allows us to formally set as $S_r(a,\delta | {\bf Y}) = -\infty$.

We next consider the case of 
${\cal L}_0(a) \neq \emptyset$.
It is easy to see that letting $n \to \infty$ in (\ref{solution4}) yields the following {\it non-asymptotic} equation to determine the second-order resolvability $b=b^{\ast}(a, \delta)$:
\begin{equation} \label{eq:8-1-1}
\sum_{i \in {\cal L}_0(a)} w(i) \Phi\left(\frac{b}{\sigma_i} \right) + \sum_{j \in {\cal L}_1(a)} w(j) = \frac{\delta}{2}.
\end{equation}
In order to specifically compute the $b=b^{\ast}(a, \delta)$, we need several classifications as in the proof of Theorem \ref{theo1}, where equation (\ref{eq:8-1-1}) implies that $a$ given $\delta$ is specified by the following conditions: If ${\cal L}_0(a) = \emptyset$ then $\sum_{j \in {\cal L}_1(a)} w(j)=\frac{\delta}{2}$ $(b = -\infty)$; If ${\cal L}_0(a) \neq \emptyset$ then
\[
\sum_{j \in {\cal L}_1(a)} w(j) < \frac{\delta}{2},
\]
\[
\sum_{i \in {\cal L}_0(a)} w(i) + \sum_{j \in {\cal L}_1(a)} w(j) \geq \frac{\delta}{2}.
\]
This is illustrated in Fig. \ref{Countably Infinite}.
\begin{figure}[h]
\begin{center}
\includegraphics[width=4.2in]{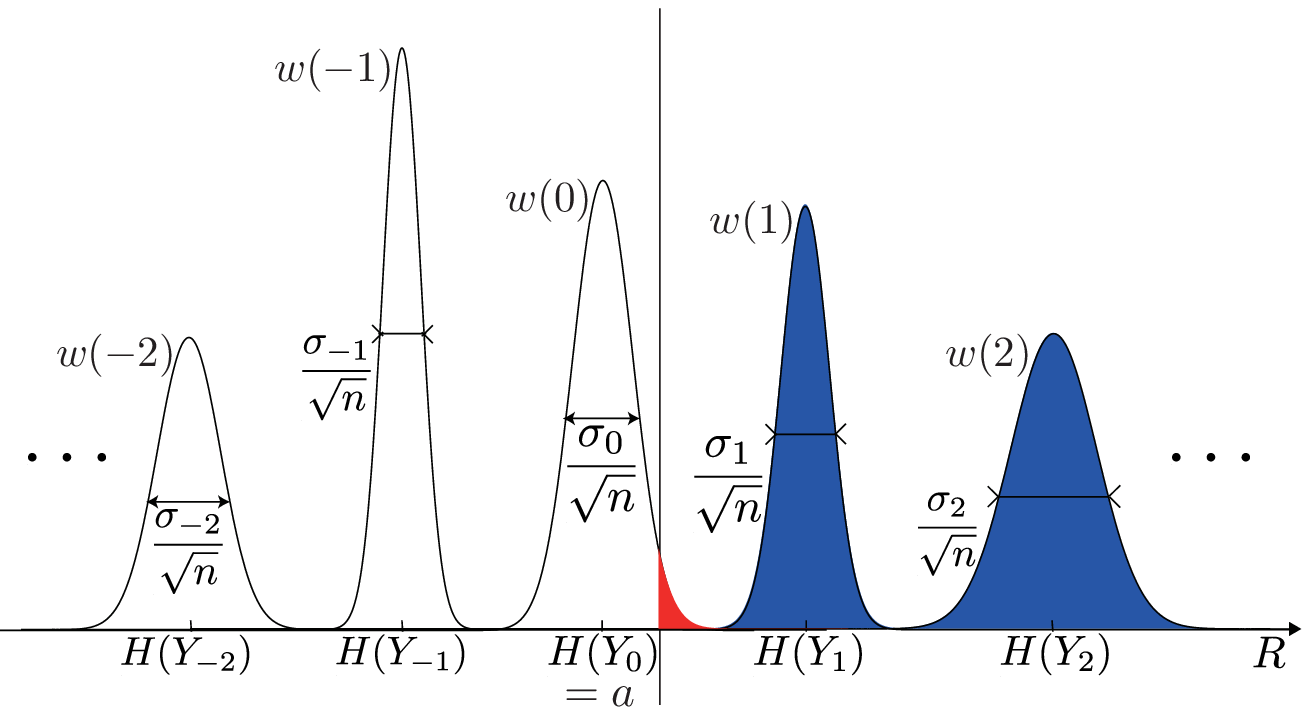}
% where an .eps filename suffix will be assumed under latex, 
% and a .pdf suffix will be assumed for pdflatex; or what has been declared
% via \DeclareGraphicsExtensions.
\centering \caption{Countably Infinite i.i.d. Sources}
\label{Countably Infinite}
\end{center}
\end{figure}

It is easy to see that these conditions yield Cases I, II, III in Theorem \ref{theo1} as a special case.
\end{IEEEproof}
\subsection{$(a,\delta)$-Resolvability for a Mixture of Countably Infinite Markovian Sources}
Next, let us consider a mixed source consisting of {\it countably infinite} Markovian sources with {\it finite} source alphabet ${\bf Y}_i = \{Y_i^n \}_{n=1}^{\infty}$, where $i \in \mathbb{Z} = \{0, \pm1, \pm2, \cdots \}$.
%The mixed source that we consider in this subsection is defined by
The mixed source that we consider in this subsection is defined by
\[
P_{Y^n}({\bf y}) = \sum_{i=-\infty}^\infty w(i) P_{Y^n_i}({\bf y}),
\]
where $w(i)$ are constants satisfying $\sum_{i = -\infty}^\infty w(i) =1$ and $w(i) \geq 0$ for all $i \in \mathbb{Z}$ and each ${\bf Y}_i = \{Y_i^n \}_{n=1}^\infty$ denotes the stationary ergodic Markovian source with irreducible transition probability $Q_i(k|j)$ $((j,k) \in {\cal Y}^2)$, where $j$ and $k$ denote consecutive symbols in the Markov transition. 
We let $\pi_i(\cdot)$ denote the stationary distribution of ${\bf Y}_i$.
The probability of a sequence ${\bf y} = y_1 y_2 \cdots y_n \in {\cal Y}^n$ emitted from Markovian source ${\bf Y}_i$ is given by
\[
P_{Y_i^n}({\bf y}) = \pi_i(y_1) \prod_{j=1}^{n-1} Q_i(y_{j+1}|y_j).
\]

The entropy rate $H(Q_i)\  (i \in \mathbb{Z})$ and the variance $\sigma_i^2 \ (i \in \mathbb{Z})$ of Markovian source ${\bf Y}_i$ are defined by
\[
H(Q_i) \equiv \sum_{(k,j) \in {\cal Y}^2} \pi_i(j)Q_i(k|j) \log \frac{1}{Q_i(k|j)},
\]
and
\begin{align*}
\sigma_i^2 \equiv & \sum_{(k,j) \in {\cal Y}^2} \pi_i(j)Q_i(k|j) \left( \log \frac{1}{Q_i(k|j)} - H(Q_i) \right)^2 \\
& + 2\sum_{(l,k,j) \in {\cal Y}^3}\pi_i(j)Q_i(l|k)Q_i(k|j) \left( \log \frac{1}{Q_i(l|k)} - H(Q_i) \right) \left(\log \frac{1}{Q_i(k|j)} - H(Q_i) \right),
\end{align*}
respectively (see Hayashi \cite[VII]{Hayashi}).
In addition, we assume that all these variances are finite, and it is also assumed that
\[
\cdots \leq H(Q_{-2}) \leq H(Q_{-1}) \leq H(Q_0) \leq H(Q_{1}) \leq H(Q_2) \leq \cdots.
\]

With this definition of mixed sources, we now have the following second-order resolvability theorem:
\begin{theorem} \label{theo:8-2}
For a mixture ${\bf Y}$ of {\it countably infinite} Markovian sources ${\bf Y}_i\ ( i \in {\mathbb{Z}})$ with {\it finite} source alphabet, the optimal size rate 
$
R_n^{\ast} =\frac{1}{n} \log M^{\ast}_n
$
is given as the solution of the asymptotic equation:
\begin{align*} 
\sum_{i=-\infty}^{\infty} w(i)\Phi \left( \frac{\sqrt{n}(R_n - H(Q_i))}{\sigma_i} \right)= \frac{\delta}{2};
\end{align*}
and, furthermore, the second-order resolvability $S_r(a,\delta|{\bf Y})$
is given as the solution $b = b^{\ast}(a, \delta)$ of the equation
\begin{align*} 
\sum_{i=-\infty}^{\infty} w(i)\Phi \left( \frac{\sqrt{n}(a - H(Q_i))+ b + o(1)}{\sigma_i} \right)= \frac{\delta}{2}.
\end{align*}
\end{theorem}
\begin{IEEEproof}
Since we assume that the source alphabet ${\cal Y}$ is finite, the number of states of this Markov chain is finite.
Then, the asymptotic normality also holds due to the central limit theorem for Markov chains (see, e.g., Chung \cite{Chung}, Billingsley \cite{Billingsley}).
Therefore, it suffices to proceed  in parallel with the arguments as made in the proof of Theorem \ref{theo:8-1} and so we omit the details.
\end{IEEEproof}
\begin{remark}
As shown in this section, our analysis for mixed sources is on the basis of the asymptotic normality of self-information. This means that the similar argument is valid for any mixture of countably infinite sources for which the asymptotic normality of self-information holds for each of the component sources. Generally speaking, one of the conditions to ensure the central limit theorem for weakly dependent random variables is, e.g., the {\it mixing} condition (see, Billingsley \cite{Billingsley}, Shields \cite{Shields}).
\end{remark}
%
%
% subsection8-3
%
%
\subsection{$(a,\delta)$-Resolvability for Sources with General Mixture}
In this subsection we consider a possible extension of Theorems \ref{theo:8-1} and \ref{theo:8-2} to the case with {\it general mixture} instead of countably infinite mixture.
It suffices here only to focus on the extension of Theorem \ref{theo:8-1}, because the basic logics underlying both extensions are the same.
A mixed source ${\bf Y} = \{ Y^n \}_{n=1}^\infty$ with {\it general mixture} is defined by
\begin{equation}
P_{Y^n}({\bf y}) = \int_{\Lambda} P_{Y_{\theta}^n}({\bf y}) dw(\theta),
\end{equation}
where $w(\theta)$ is an arbitrary probability measure on the parameter space $\Lambda$, and ${\bf Y}_\theta = \{ Y_{\theta}^n \}_{n=1}^\infty \ (\theta \in \Lambda)$ are i.i.d. sources with finite alphabet. 

With this definition, we have the following formal extension of Theorem \ref{theo:8-1}:
\begin{theorem} \label{theo:8-3}
For a source ${\bf Y}$ with general mixture $w(\theta)$ of i.i.d. sources ${\bf Y}_\theta$ with finite alphabet, the optimal size rate 
$
R_n^{\ast} =\frac{1}{n} \log M^{\ast}_n
$
is given as the solution for $R_n$ of the asymptotic equation:
\begin{align*} 
\int_{\Lambda} \Phi \left( \frac{\sqrt{n}(R_n - H(Y_\theta))}{\sigma_\theta} \right) d w(\theta) = \frac{\delta}{2};
\end{align*}
and, furthermore, the second-order resolvability $S_r(a,\delta|{\bf Y})$
is given as the solution $b = b^{\ast}(a, \delta)$ of the asymptotic equation
\begin{align*} 
\int_{\Lambda} \Phi \left( \frac{\sqrt{n}(a - H(Y_\theta))+ b + o(1)}{\sigma_\theta} \right) dw(\theta)= \frac{\delta}{2},
\end{align*}
where $\sigma^2_\theta$ is the variance of $\log \frac{1}{P_{Y_{\theta}}(Y_{\theta})}$.
\end{theorem}
\begin{IEEEproof}
The second inequality of Lemma \ref{lemma} is not necessarily valid in the case with general mixture. Nevertheless, we can slightly modify it so as to be applicable to this general case. As for the details, see the proof of Han \cite[Lemma 1.4.4]{Han}.
\end{IEEEproof}
\begin{remark}
Let us define, instead of ${\cal L}_0(a)$ and ${\cal L}_1(a)$ as above, the subsets $\Lambda_0(a)$ and $\Lambda_1(a)$ of $\Lambda$ as follows:
\[
{\Lambda}_0(a) = \{ \theta \in \Lambda | {H}(Y_\theta) = a\},
\]
\[
{\Lambda}_1(a) = \{ \theta \in \Lambda | {H}(Y_\theta) > a \}.
\]
Then, the equation (\ref{eq:8-1-1}) to determine the $(a, \delta)$-resolvability $b = b^{\ast}(a, \delta)$ is extended accordingly to the case with general mixture $w(\theta)$ as follows:
\begin{equation} 
\int_{\Lambda_0(a)} \Phi\left(\frac{b}{\sigma_\theta} \right) dw(\theta) + \int_{\Lambda_1(a)} dw(\theta) = \frac{\delta}{2},
\end{equation}
which implies that $a$ given $\delta$ is specified by the following conditions: If $\int_{\Lambda_0(a)}dw(\theta) = 0$ then $\int_{\Lambda_1(a)}dw(\theta) = \frac{\delta}{2}$ $(b = -\infty)$; If $\int_{\Lambda_0(a)}dw(\theta) > 0$ then
\[
\int_{\Lambda_1(a)} dw(\theta) < \frac{\delta}{2},
\]
\[
\int_{\Lambda_0(a)} dw(\theta) + \int_{\Lambda_1(a)} dw(\theta) \geq \frac{\delta}{2}.
\]
\end{remark}

\section{Concluding Remarks}
We have so far considered the second-order achievability to evaluate the finer structure of random number generation for mixed sources.
The class of mixed sources is very important, because all of stationary sources can be regarded as forming mixed sources obtained by mixing stationary ergodic sources with respect to appropriate probability measures.
Although, in general, mixed sources do not have the asymptotic normality.
So, our result is also meaningful, we have demonstrated that the analysis based on the {\it two-peak} asymptotic normality is still effective also for sources whose self-information spectrum does {\it not} have a single asymptotic normality.

As shown in the proofs of the present paper, the information spectrum approach is substantial in the analysis of the {\it second-order} achievable rates.
In particular, Lemma \ref{lemma} is a simple but enables us to work with the {\it multi-peak} asymptotic normality for mixed sources (cf. the proof of Theorem \ref{theo:8-1}).

Polyanskiy, Poor and Verd\'{u} \cite{Polyanskiy2011} has derived the {\it second-order} capacity in either {\it ergodic} or {\it nonergodic} setting for the Gilbert-Elliott channel (GEC), which consists of two {\it binary} symmetric channels. %In the GEC, the crossover probability is a Markov chain. 
Their results for the {\it nonergodic} case is also based on a mixture of two Gaussian distributions.
On the other hand, since our analysis is based on the information spectrum methods, the results are valid also for a mixture of {\it countably infinite} sources (i.i.d. with {\it countably infinite} alphabet or Markov with {\it finite} alphabet) as well as a {\it general mixture} of i.i.d. sources with {\it finite} alphabet as was shown in Section VIII.
It should be emphasized that, throughout in the paper, we have established the {\it asymptotic} or {\it nonasymptotic} equations for determining the second-order resolvability, intrinsic randomness, and fixed-length source coding rate. The forms of these equations include that as shown in \cite{Polyanskiy2011} as a special case. 
We observe that the second-order capacity of finite state Markov channels (see, \cite{Goldsmith1996}) or Fritchman channels (see, \cite{Fritchman}), which is generalizations of the GEC, can be established by means of the similar arguments. This will be reported in a forthcoming paper \cite{RHfc}.
% if have a single appendix:
%\appendix[Proof of the Zonklar Equations]
% or
%\appendix  % for no appendix heading
% do not use \section anymore after \appendix, only \section*
% is possibly needed
% use appendices with more than one appendix
% then use \section to start each appendix
% you must declare a \section before using any
% \subsection or using \label (\appendices by itself
% starts a section numbered zero.)
%
\appendices
\section{Proof of Theorem \ref{theo:2-4}}  \label{app1}
\renewcommand{\theequation}{A.\arabic{equation}}
\setcounter{equation}{0}
This appendix concerns the operational relationship between resolvability and fixed-length source coding. Although the proof literally mimics the argument given by Han \cite[p.163]{Han}, we will repeat it here for the reader's convenience.

First, we show how to construct an $(n, M_n, \varepsilon_n)$ source code, given a resolvability mapping $\phi_n: {\cal U}_{M_n} \to {\cal Y}^n$. Set $\tilde{Y}^n = \phi_n(U_{M_n})$, then we can define the subset ${\cal S}_0$ of ${\cal Y}^n$
by
\[
{\cal S}_0 = \left\{ {\bf y} \in {\cal Y}^n \left| P_{\tilde{Y}^n}({\bf y}) >0  \right. \right\}.
\]
Clearly, we have $|{\cal S}_0| \leq M_n$ which enables us to define the source encoder $\varphi_n: {\cal Y}^n \to {\cal U}_{M_n}$ which transforms each element of ${\cal S}_0$ to a distinct element of ${\cal U}_{M_n}$ and all elements of ${\cal Y}^n \setminus {\cal S}_0$ to $1$. In addition, if we define the source decoder $\psi_n$ as the inverse mapping of $\varphi_n|{}_{{\cal S}_0}$, the $(\varphi_n, \psi_n)$ becomes an $(n, M_n, \varepsilon_n)$ source code for the source ${\bf Y} = \{Y^n \}_{n=1}^\infty$ such that
\[
\varepsilon_n = \Pr \left\{ Y^n \notin {\cal S}_0 \right\}.
\]
Here, from the other definition of the variational distance:
\[
d(Z,{\tilde{Z}}) = 2\sup_{A: A \subset {\cal Z} } \left| P_{Z}(A) - P_{\tilde{Z}}(A) \right|,
\]
it follows that
\[
\Pr \left\{ \tilde{Y}^n \in {\cal S}_0 \right\} - \Pr \left\{ {Y}^n \in {\cal S}_0 \right\} \leq \frac{1}{2} d(Y^n, \tilde{Y}^n).
\]
Considering that $\Pr \left\{ \tilde{Y}^n \in {\cal S}_0 \right\} = 1$ holds, we see that
\[
\Pr \left\{ {Y}^n \notin {\cal S}_0 \right\} \leq \frac{1}{2} d(Y^n, \tilde{Y}^n).
\]
Consequently, we obtain
\begin{equation} \label{eq:c-1}
\varepsilon_n \leq \frac{1}{2} d(Y^n, \tilde{Y}^n).
\end{equation}
In this way we can construct an $(n, M_n, \varepsilon_n)$ source code satisfying (\ref{eq:c-1}) with the same $M_n$ as in the resolvability mapping $\phi_n: {\cal U}_{M_n} \to {\cal Y}^n$, which implies that if $R > S_r(a, \delta|{\bf Y})$ then $R > L_f(a, \frac{\delta}{2}|{\bf Y})$.

Next, let us show how to construct a resolvability mapping $\phi_n$,
%$U_{M_n}$ of size $M_n$ 
given an $(n, M_n, \varepsilon_n)$ source code.
We can define the subset ${\cal S}_0$ of ${\cal Y}^n$ by
\[
{\cal S}_0 = \left\{ {\bf y} \in {\cal Y}^n \left| {\bf y} = \psi_n(\varphi_n({\bf y})) \right.  \right\},
\]
where $(\varphi_n, \psi_n)$ denotes the pair of encoder and decoder of the $(n, M_n, \varepsilon_n)$ code.
For an arbitrarily small $\gamma >0$ set $M'_n$ = $M_n e^{\sqrt{n}\gamma}$. From the same argument as in the proof of Lemma \ref{lemma1} (cf. Han \cite{Han}) with $U_{M'_n}$, ${\cal S}_0$ and ${\cal U}_{M'_n}$ in place of $X^n$, $T_n(z_n)$ and $S_n(z_n + \gamma)$, we can construct a mapping $\phi_n: {\cal U}_{M'_n} \to {\cal Y}^n$ such that
\begin{equation} \label{eq;c-2}
d(Y^n, \tilde{Y}^n) \leq 2 \Pr\left\{ Y^n \notin {\cal S}_0  \right\} + 2 e^{-\sqrt{n}\gamma},
\end{equation}
where $\tilde{Y}^n = \phi_n(U_{M'_n})$.
Since $\varepsilon_n = \Pr \{ Y^n \notin {\cal S}_0 \}$, (\ref{eq;c-2}) can be expressed as
\begin{equation*} \label{eq:app3-1}
d(Y^n, \tilde{Y}^n) \leq 2 \varepsilon_n + 2 e^{-\sqrt{n}\gamma}.
\end{equation*}
In this way we can construct the resolvability mapping $\tilde{Y}^n = \phi_n(U_{M'_n})$, which implies that if $R > L_f(a, \varepsilon|{\bf Y})$ then $R + \gamma > S_r(a, 2\varepsilon|{\bf Y})$.
\section{Proof of Lemma \ref{lemma}} \label{app2}
Although the proof to be shown below is implicitly contained in Han \cite{Han}, we explicitly summarize it here for the reader's convenience.
At first we show the first inequality. Set a sequence $\{\gamma_n \}_{n=1}^{\infty}$  satisfying $\gamma_1 > \gamma_2 > \cdots > 0,$ $\gamma_n \to 0$, $\sqrt{n} \gamma_n \to \infty$. Then it holds that
\begin{eqnarray*}
\lefteqn{\Pr \left\{ \frac{-\!\log P_{Y^n}(Y^n_{i}) }{\sqrt{n}}  - \frac{-\!\log P_{Y^n_{i}}(Y^n_{i}) }{\sqrt{n}} \leq -\gamma_n \right\}} \\
& = & \sum_{{\bf y} \in D_{n}(i)} P_{Y^n_i}({\bf y}) \\
& \leq & \sum_{{\bf y} \in D_n(i)} P_{Y^n}({\bf y}) e^{- \sqrt{n}\gamma_n} \leq e^{- \sqrt{n}\gamma_n},
\end{eqnarray*}
for $i = 1,2$, where
\begin{eqnarray*}
%\lefteqn{D_n(i)} \\
%& = & \left\{{\bf y}\!\in\!{\cal Y}^n \left| \frac{-\!\log P_{Y^n}({\bf y}) }{\sqrt{n}}\!-\!\frac{-\!\log P_{Y^n_{i}}({\bf y}) }{\sqrt{n}}\!\leq - \gamma_n \right. \right\}.
D_n(i) =  \left\{{\bf y}\!\in\!{\cal Y}^n \left| \frac{-\!\log P_{Y^n}({\bf y}) }{\sqrt{n}}\!-\!\frac{-\!\log P_{Y^n_{i}}({\bf y}) }{\sqrt{n}}\!\leq - \gamma_n \right. \right\}.
\end{eqnarray*}
This means that 
%\begin{multline*}
%\Pr \left\{ \frac{-\!\log P_{Y^n_{i}}(Y^n_{i}) }{\sqrt{n}} - \gamma_n < \frac{-\!\log P_{Y^n}(Y^n_{i}) }{\sqrt{n}} \right\} \\
% \geq  1 - e^{- \sqrt{n}\gamma_n},
%\end{multline*}
\begin{equation*}
\Pr \left\{ \frac{-\!\log P_{Y^n_{i}}(Y^n_{i}) }{\sqrt{n}} - \gamma_n < \frac{-\!\log P_{Y^n}(Y^n_{i}) }{\sqrt{n}} \right\} \geq  1 - e^{- \sqrt{n}\gamma_n},
\end{equation*}
holds for $ i = 1,2$.
So we have
\[
\frac{-\!\log P_{Y^n_{i}}(Y^n_{i}) }{\sqrt{n}} -  \gamma_n < \frac{-\!\log P_{Y^n}(Y^n_{i}) }{\sqrt{n}}
\]
with probability $1 - e^{- \sqrt{n}\gamma_n}$.
So, we have for $i = 1,2$
\begin{eqnarray*}
%\lefteqn{ \Pr\left\{ \frac{-\!\log P_{Y^n}(Y^n_{i}) }{\sqrt{n}}\!\geq\! z_n \right\}} \nonumber \\
%&\!\geq\!&  \Pr\left\{ \frac{-\!\log P_{Y^n_{i}}(Y^n_{i}) }{\sqrt{n}} -  \gamma_n \geq z_n \right\} - e^{- \sqrt{n}\gamma_n},
\Pr\left\{ \frac{-\!\log P_{Y^n}(Y^n_{i}) }{\sqrt{n}}\!\geq\! z_n \right\} \geq \Pr\left\{ \frac{-\!\log P_{Y^n_{i}}(Y^n_{i}) }{\sqrt{n}} -  \gamma_n \geq z_n \right\} - e^{- \sqrt{n}\gamma_n},
\end{eqnarray*}
which is the first inequality of the lemma.
Secondly, we show the second inequality of the lemma.
Set 
\[
S_n(z_n) = \left\{ {\bf y} \in {\cal Y}^n\left| P_{Y^n}({\bf y}) \leq e^{ - \sqrt{n}z_n} \right. \right\}
\]
\[
S_n^{(i)}(z_n) = \left\{ {\bf y} \in {\cal Y}^n\left| P_{Y^n_{i}}({\bf y}) \leq \frac{ e^{ - \sqrt{n}z_n} }{w(i)} \right. \right\}
\]
From the property of mixed sources, ${\bf y} \in S_n^{(i)}(z_n)$ ($i = 1,2$) holds for ${\bf y} \in S_n(z_n)$. 
This means that 
\[
S_n(z_n) \subset S_n^{(i)}(z_n)
\]
($i = 1,2$).
Moreover, since, by assumption, $\gamma_n \geq \frac{-\log w(i)}{\sqrt{n}}$ ($i =1,2$) hold for sufficiently large $n$,  we have
\begin{eqnarray*}
\lefteqn{ \Pr \left\{ \frac{-\log P_{Y^n}(Y^n_{i}) }{\sqrt{n}}\!\geq\! z_n \right\}} \nonumber \\
& = & \Pr\left\{  P_{Y^n}(Y^n_{i}) \leq e^{ - \sqrt{n}z_n  } \right\}\\
& = & P_{Y^n_{i}}(S_n(z_n)) \\
&\!\leq\!&  P_{Y^n_{i}}( S_n^{(i)}(z_n) ) \\
&\!=\!&  \Pr\left\{ \frac{-\!\log P_{Y^n_{i}}(Y^n_{i}) }{\sqrt{n}}\!\geq\! z_n \!-\!\frac{-\!\log w(i)}{\sqrt{n}} \right\} \\
&\!\leq\!& \Pr\left\{ \frac{-\!\log P_{Y^n_{i}}(Y^n_{i}) }{\sqrt{n}}\!\geq\! z_n \!-\!\gamma_n \right\},
\end{eqnarray*}
for sufficiently large $n$, which is the second inequality.
\IEEEQED
%
%   appendix c
%
\section{Proof of Theorem \ref{theo1}} \label{app4}
\renewcommand{\theequation}{C.\arabic{equation}}
\setcounter{equation}{0}
\begin{IEEEproof}[Proof of Case I] 

A simplest way to prove Theorem \ref{theo1} is to first apply Theorem \ref{theo:2-1} to the present case of mixed sources and to proceed to the computation of necessary quantities.
Here, however, more basically we start along with Lemma \ref{lemma1} and Lemma \ref{lemma2} in order to reveal the fundamental logic underlying the whole process of random number generation.
Actually, the computations needed in the first way of proof are contained in those needed in the second way of proof; more exactly, both computations are substantially the same.

We show the proof of { Case I} by using Lemma \ref{lemma1} and Lemma \ref{lemma2}. The proof consists of two parts.
\vspace*{\baselineskip}

{\it 1) Direct Part}: 

Set $M_n = e^{nH(Y_1)+T_1 \sqrt{n} + \gamma }$, where $\gamma > 0$ is an arbitrarily small number. Then, trivially it holds that
$$
\limsup_{n \rightarrow \infty} \frac{\log M_n - nH(Y_1)}{\sqrt{n}} \leq T_1.
$$
Thus, it is enough to show that there exists a mapping $\phi_n$ such that
$$
\limsup_{n \rightarrow \infty} d(\phi_n(U_{M_n}),Y^n) \leq \delta.
$$
%then Direct Part is proved.

On the other hand, set $z_n = \frac{nH(Y_1)+T_1\sqrt{n} }{\sqrt{n}} - \gamma$, then $z_n + \gamma \leq \frac{\log M_n}{\sqrt{n}}$ holds. Thus, from Lemma \ref{lemma1} and (\ref{reduced}) there exists a mapping $\phi_n$ such that
\begin{eqnarray*}
\lefteqn{\frac{1}{2}d(\phi_n(U_{M_n}),Y^n)} \\
%&\!\leq\!& \Pr\left\{ \frac{nH(Y_1)+T_1\sqrt{n}}{\sqrt{n}} - \gamma < \frac{1}{\sqrt{n}}\log \frac{1}{P_{Y^n}(Y^n)}  \right\} \\
%&      & + e^{- \sqrt{n}\gamma}.% \\
&\!\leq\!& \Pr\left\{ \frac{nH(Y_1)+T_1\sqrt{n}}{\sqrt{n}} - \gamma < \frac{1}{\sqrt{n}}\log \frac{1}{P_{Y^n}(Y^n)}  \right\} + e^{- \sqrt{n}\gamma}.% \\
\end{eqnarray*}
Moreover, from Lemma \ref{lemma}, there exists a mapping $\phi_n$ such that
\begin{align*} \label{a1}
\lefteqn{\limsup_{n \to \infty}\frac{1}{2}d(\phi_n(U_{M_n}),Y^n)} \nonumber \\
& \leq \limsup_{n \to \infty} \Pr\left\{ \frac{nH(Y_1)\!+\!T_1\sqrt{n}}{\sqrt{n}}\!-\!\gamma \!<\!\frac{1}{\sqrt{n}}\log \frac{1}{P_{Y^n}(Y^n)}  \right\}  \nonumber \\
%& = \limsup_{n \to \infty} \sum_{i=1}^2 \Pr\left\{ \frac{nH(Y_1)\!+\!T_1\sqrt{n}}{\sqrt{n}} - \gamma \right. \nonumber \\
%&  \hspace*{1.5cm} \left. <\!\frac{1}{\sqrt{n}}\log \frac{1}{P_{Y^n}(Y_i^n)}  \right\} w(i)  \nonumber \\
& = \limsup_{n \to \infty} \sum_{i=1}^2 \Pr\left\{ \frac{nH(Y_1)\!+\!T_1\sqrt{n}}{\sqrt{n}} - \gamma  <\!\frac{1}{\sqrt{n}}\log \frac{1}{P_{Y^n}(Y_i^n)}  \right\} w(i)  \nonumber \\
%& \leq \sum_{i=1}^2  \limsup_{n \to \infty}  \Pr\left\{ \frac{nH(Y_1)+T_1\sqrt{n}}{\sqrt{n}} - \gamma -\gamma_n \right.  \\
%& \hspace*{1.5cm}   \left.  < \frac{1}{\sqrt{n}}\log \frac{1}{P_{Y_i^n}(Y_i^n)}  \right\} w(i) \nonumber \\
& \leq \sum_{i=1}^2  \limsup_{n \to \infty}  \Pr\left\{ \frac{nH(Y_1)+T_1\sqrt{n}}{\sqrt{n}} - \gamma -\gamma_n < \frac{1}{\sqrt{n}}\log \frac{1}{P_{Y_i^n}(Y_i^n)}  \right\} w(i) \nonumber \\
%& \leq \sum_{i=1}^2 \limsup_{n \to \infty} \Pr\left\{\frac{- \log P_{Y_i^n}(Y_i^n)\!-\!nH(Y_1)}{\sqrt{n}}\!>\!T_1\!-\! 2\gamma\right\}  \\
%&  \ \ \ \cdot w(i) \\
& \leq \sum_{i=1}^2 \limsup_{n \to \infty} \Pr\left\{\frac{- \log P_{Y_i^n}(Y_i^n)\!-\!nH(Y_1)}{\sqrt{n}}\!>\!T_1\!-\! 2\gamma\right\} w(i),
%& =  \sum_{i=1}^2 \limsup_{n \to \infty} \Pr\left\{\frac{\!-\!\log P_{Y_i^n}(Y_i^n)\!-\!nH(Y_1)}{\sqrt{n} \sigma_i} > \frac{T_1\!-\!%2\gamma}{\sigma_i} \right\}  \\
%&  \ \ \ w(i),
%& =  \sum_{i=1}^2 \limsup_{n \to \infty} \Pr\left\{\frac{\!-\!\log P_{Y_i^n}(Y_i^n)\!-\!nH(Y_1)}{\sqrt{n} \sigma_i} > \frac{T_1\!-\!2\gamma}{\sigma_i} \right\} w(i),
\end{align*}
because $\gamma_n > 0$ is specified in Lemma \ref{lemma}, and so $\gamma_n < \gamma$ holds for sufficiently large $n$.
Then, noting that $H(Y_1)=H(Y_2)$ holds,  we have from the asymptotic normality (by virtue of the central limit theorem)
\begin{align*}
\lefteqn{\limsup_{n \rightarrow \infty}\Pr\left\{  \frac{\!-\log P_{Y_i^n}(Y_i^n)- nH(Y_1) }{\sqrt{n} \sigma_i} > \frac{T_1-2\gamma}{\sigma_i} \right\}} \\
& = \int^{\infty}_{\frac{T_1 -2\gamma}{\sigma_i}} \frac{1}{\sqrt{2\pi}}\exp\left[ -\frac{z^2}{2} \right] dz \\
& =  \int^{\infty}_{\frac{T_1}{\sigma_i}} \frac{1}{\sqrt{2\pi}}\exp\left[\!-\!\frac{z^2}{2} \right] dz \!+\! \int_{\frac{T_1\!-\!2\gamma }{\sigma_i}}^{\frac{T_1}{\sigma_i}} \frac{1}{\sqrt{2\pi}} \exp  \left[ \!-\!\frac{z^2}{2} \right] dz.
\end{align*} 
Here, by the continuity of the normal distribution, 
$$
 \int_{\frac{T_1\!-\!2\gamma }{\sigma_i}}^{\frac{T_1}{\sigma_i}} \frac{1}{\sqrt{2\pi}} \exp  \left[ \!-\!\frac{z^2}{2} \right] dz \to 0
$$
as $\gamma \to 0$.
Thus, noting that $\gamma > 0$ is an arbitrarily small, we have
\begin{align*} 
\lefteqn{\limsup_{n \to \infty}\frac{1}{2}d(\phi_n(U_{M_n}),Y^n)} \\
& \leq  \sum_{i=1}^2 w(i) \int^{\infty}_{\frac{T_1}{\sigma_i}} \frac{1}{\sqrt{2\pi}}\exp\left[ -\frac{z^2}{2} \right] dz  = \frac{\delta}{2} .
\end{align*}
Therefore, Direct Part has been proved.
%
%
%
% latter half
%
%

\vspace*{\baselineskip}

{\it 2) Converse Part}: 

We consider a constant $T_1' < T_1$.
Assuming that $T_1'$ is $(H(Y_1),\delta)$-achievable, we shall show a contradiction.
Since we assume that $T_1'$ is $(H(Y_1),\delta)$-achievable, there exists a mapping $\phi_n$ such that
$$
\limsup_{n \rightarrow \infty} d(\phi_n(U_{M_n}),Y^n) \leq \delta,
$$
\begin{equation*} \label{C1}
\limsup_{n \rightarrow \infty}  \frac{\log M_n - n H(Y_1) }{\sqrt{n}}  \leq T_1',
\end{equation*}
which means that there exists a constant $\gamma > 0$ satisfying
$$
\frac{\log M_n - n H(Y_1) }{\sqrt{n}}  < T_1'+ \gamma < T_1,
$$
for sufficiently large $n$.

On the other hand, set $z_n = \frac{nH(Y_1)+T'_1\sqrt{n}}{\sqrt{n}} + \gamma $. Then $z_n > \frac{\log M_n}{\sqrt{n}}$ holds.
Thus, from Lemma \ref{lemma2} and (\ref{reduced}), for any mapping $\phi_n$ it holds that
\begin{eqnarray*}
\lefteqn{\frac{1}{2}d(\phi_n(U_{M_n}),Y^n)} \\
%& \geq & \Pr\left\{\frac{nH(Y_1)+T'_1\sqrt{n}}{\sqrt{n}} + 2\gamma <  \frac{1}{\sqrt{n}}\log \frac{1}{P_{Y^n}(Y^n)} \right\}  \\
%& & - e^{- \sqrt{n} \gamma}.
& \geq & \Pr\left\{\frac{nH(Y_1)+T'_1\sqrt{n}}{\sqrt{n}} + 2\gamma <  \frac{1}{\sqrt{n}}\log \frac{1}{P_{Y^n}(Y^n)} \right\} - e^{- \sqrt{n} \gamma}.
\end{eqnarray*}
Thus, from Lemma \ref{lemma}, for any mapping $\phi_n$ we have
\begin{align} \label{C2}
\lefteqn{\liminf_{n \to \infty} \frac{1}{2}d(\phi_n(U_{M_n}),Y^n)} \nonumber \\
&\geq \liminf_{n \to \infty} \Pr\left\{\frac{nH(Y_1)\!+\!T'_1\sqrt{n}}{\sqrt{n}}\!+\!2\gamma <  \frac{1}{\sqrt{n}}\log \frac{1}{P_{Y^n}(Y^n)} \right\} \nonumber \\
%&\!=\!\liminf_{n \to \infty} \sum_{i=1}^2 \Pr\left\{\frac{nH(Y_1)\!+\!T'_1\sqrt{n}}{\sqrt{n}}\!+\!2\gamma\!< \!\frac{1}{\sqrt{n}}\log \frac{1}{P_{Y^n}(Y_i^n)} \right\} \nonumber \\
%& \ \ \ \cdot w(i) \nonumber \\
&\!=\!\liminf_{n \to \infty} \sum_{i=1}^2 \Pr\left\{\frac{nH(Y_1)\!+\!T'_1\sqrt{n}}{\sqrt{n}}\!+\!2\gamma\!< \!\frac{1}{\sqrt{n}}\log \frac{1}{P_{Y^n}(Y_i^n)} \right\} w(i) \nonumber \\
%&\geq \sum_{i=1}^2 \liminf_{n \to \infty} \Pr\left\{\frac{nH(Y_1)+T'_1\sqrt{n}}{\sqrt{n} } + 2\gamma + \gamma_n  \right. \nonumber \\
%& \hspace*{1cm} \left. <  \frac{1}{\sqrt{n}}\log \frac{1}{P_{Y_i^n}(Y_i^n)}  \right\} w(i) \nonumber \\ 
&\geq \sum_{i=1}^2 \liminf_{n \to \infty} \Pr\left\{\frac{nH(Y_1)+T'_1\sqrt{n}}{\sqrt{n} } + 2\gamma + \gamma_n  <  \frac{1}{\sqrt{n}}\log \frac{1}{P_{Y_i^n}(Y_i^n)}  \right\} w(i) \nonumber \\ 
%& = \sum_{i=1}^2 \liminf_{n \to \infty} \Pr\left\{\frac{- \log P_{Y_i^n}(Y_i^n) \!-\! nH(Y_1)}{\sqrt{n}}\!>\!T'_1\!+\!2\gamma\!+\!\gamma_n \right\} \nonumber \\
%& \ \ \ \cdot w(i) \nonumber \\
& = \sum_{i=1}^2 \liminf_{n \to \infty} \Pr\left\{\frac{- \log P_{Y_i^n}(Y_i^n) \!-\! nH(Y_1)}{\sqrt{n}}\!>\!T'_1\!+\!2\gamma\!+\!\gamma_n \right\}  w(i) \nonumber \\
%& \geq \sum_{i=1}^2 \liminf_{n \to \infty} \Pr\left\{\frac{- \log P_{Y_i^n}(Y_i^n) - nH(Y_1)}{\sqrt{n}} > T'_1 + 3\gamma \right\} \nonumber \\
%& \ \ \ \cdot w(i),
& \geq \sum_{i=1}^2 \liminf_{n \to \infty} \Pr\left\{\frac{- \log P_{Y_i^n}(Y_i^n) - nH(Y_1)}{\sqrt{n}} > T'_1 + 3\gamma \right\} w(i),
\end{align}
because $\gamma > \gamma_n$ holds for sufficiently large $n$ (cf. Lemma \ref{lemma}).
%Since $T_1' + \gamma < T_1$ holds, we have
%\begin{eqnarray*} 
%\lefteqn{\sum_{i=1}^2 \liminf_{n \to \infty} \Pr\left\{\frac{- \log P_{Y_i^n}(Y_i^n) - nH(Y_1)}{\sqrt{n}} > T'_1 + \gamma \right\}w(i)} \\
%& > & \sum_{i=1}^2 \liminf_{n \to \infty} \Pr\left\{\frac{- \log P_{Y_i^n}(Y_i^n) - nH(Y_1)}{\sqrt{n}} > T_1 \right\} \\
%&    & \cdot w(i),
%\end{eqnarray*}
%where the above strict inequality follows from the asymptotic normality.
Then, by virtue of the asymptotic normality, for $i =1,2$ it holds that
\begin{align*}
&\liminf_{n \to \infty} \Pr\left\{\frac{-\log P_{Y_i^n}(Y_i^n) - nH(Y_1)}{\sqrt{n}\sigma_i} > \frac{T'_1+3\gamma}{\sigma_i} \right\} \\
& =  \int_{\frac{T'_1+3\gamma}{\sigma_i} }^{\infty} \frac{1}{\sqrt{2\pi}}\exp\left[ -\frac{z^2}{2} \right] dz  \\
& >  \int_{\frac{T_1}{\sigma_i} }^{\infty} \frac{1}{\sqrt{2\pi}}\exp\left[ -\frac{z^2}{2} \right] dz,
\end{align*}
because we can let $T_1'+3\gamma < T_1$ hold by letting $\gamma \to 0$, and by substituting the above inequality into (\ref{C2}) for any mapping $\phi_n$, we have
\begin{eqnarray*} \label{C3}
\lefteqn{\liminf_{n \to \infty} \frac{1}{2}d(\phi_n(U_{M_n}),Y^n)} \\
& > & \sum_{i=1}^2 w(i) \int_{\frac{T_1}{\sigma_i} }^{\infty} \frac{1}{\sqrt{2\pi}}\exp\left[ -\frac{z^2}{2} \right] dz = \frac{\delta}{2}.
\end{eqnarray*}
%Please note that from the continuity of normal distribution $\delta_2 \to 0$ holds if $\delta \to 0$ holds.
This is a contradiction.
Therefore, the converse part has been proved. 
\end{IEEEproof}
%
%
%  Case II and Case III
%
%
\vspace*{\baselineskip}
\begin{IEEEproof}[Proofs of Case II and Case III]

The proofs of {Case II} and {Case III} are similar to the proof of {Case I}. Here, we show only differences in the proofs.

\vspace*{\baselineskip}

{\it 1) Direct Part}:

%Let $\overline{T}_2 = T_2 - 2 \gamma$, where $\gamma$ is an arbitrary small number.
a) Similarly to {\it Case I}: we have the proof of {\it Case II} as follows.
\begin{align*} 
\lefteqn{\limsup_{n \to \infty} \frac{1}{2} d \left( \phi_n(U_{M_n}),Y^n \right)} \nonumber \\ 
&\!\leq \! \sum_{i=1}^2 \limsup_{n \to \infty} \Pr\left\{ \frac{-\!\log P_{Y_i^n}(Y_i^n)\!-\!nH(Y_1)}{\sqrt{n}}\!>\!T_2\!-\!2 \gamma \right\}w(i)  \nonumber \\
&\!=\!\sum_{i=1}^2 \limsup_{n \to \infty} \Pr\left\{ \frac{\!-\!\log P_{Y_i^n}(Y_i^n)\!-\!nH(Y_1)}{\sqrt{n} \sigma_i}\!>\! \frac{T_2\!-\!2 \gamma}{\sigma_i}  \right\} w(i)  \\
%& = \limsup_{n \to \infty} \Pr\left\{ \frac{\!-\!\log P_{Y_1^n}(Y_1^n)\!-\!nH(Y_1)}{\sqrt{n} \sigma_1}\!>\! \frac{T_2\!-\!2 \gamma}{\sigma_1}  \right\} w(1) \\
%& \ \ \ \!+\!\limsup_{n \to \infty} \Pr\left\{ \frac{\!-\!\log P_{Y_2^n}(Y_2^n)\!-\!nH(Y_1)}{\sqrt{n} \sigma_2}\!>\!\frac{T_2\!-\!2 \gamma}{\sigma_2}  \right\} w(2) \\
& = \limsup_{n \to \infty} \Pr\left\{ \frac{\!-\!\log P_{Y_1^n}(Y_1^n)\!-\!nH(Y_1)}{\sqrt{n} \sigma_1} > \frac{T_2\!-\!2 \gamma}{\sigma_1}  \right\} w(1) \\
%& \ \ \ + \limsup_{n \to \infty} \Pr\left\{ \frac{\!-\!\log P_{Y_2^n}(Y_2^n)\!-\!nH(Y_2)}{\sqrt{n} \sigma_2} \right. \\
%& \ \ \ \ \ \left. > \frac{n\left( H(Y_1)- H(Y_2)\right) }{\sqrt{n} \sigma_2} + \frac{T_2 - 2 \gamma}{\sigma_2}  \right\} w(2).
& \ \ \ + \limsup_{n \to \infty} \Pr\left\{ \frac{\!-\!\log P_{Y_2^n}(Y_2^n)\!-\!nH(Y_2)}{\sqrt{n} \sigma_2} > \frac{n\left( H(Y_1)- H(Y_2)\right) }{\sqrt{n} \sigma_2} + \frac{T_2 - 2 \gamma}{\sigma_2}  \right\} w(2).
\end{align*} 
Here, note that $H(Y_1) > H(Y_2)$ holds. 
This means that for any constant $W_1 > 0$
\[
\frac{n\left( H(Y_1)- H(Y_2)\right) }{\sqrt{n} \sigma_2} + \frac{T_2 - 2 \gamma}{\sigma_2} > W_1,
\]
holds for sufficiently large $n$. Thus, taking account of $H(Y_1) > H(Y_2)$, we have
\begin{IEEEeqnarray*}{rLl}
%\lefteqn{\limsup_{n \to \infty} \Pr\left\{ \frac{\!-\!\log P_{Y_2^n}(Y_2^n)\!-\!nH(Y_2)}{\sqrt{n} \sigma_2} \right. } \\
%& \ \ \ \ \ \  \left. > \frac{n\left( H(Y_1)- H(Y_2)\right) }{\sqrt{n} \sigma_2} + \frac{T_2 - 2 \gamma}{\sigma_2}  \right\} \\
\lefteqn{\limsup_{n \to \infty} \Pr\left\{ \frac{\!-\!\log P_{Y_2^n}(Y_2^n)\!-\!nH(Y_2)}{\sqrt{n} \sigma_2} > \frac{n\left( H(Y_1)- H(Y_2)\right) }{\sqrt{n} \sigma_2} + \frac{T_2 - 2 \gamma}{\sigma_2}  \right\} } \\
& \leq &  \limsup_{n \to \infty} \Pr\left\{ \frac{-\!\log P_{Y^n_{2}}(Y^n_{2}) - nH(Y_2)}{\sqrt{n} \sigma_2} > W_1 \right\} \\
& = & \int_{W_1}^{\infty} \frac{1}{\sqrt{2\pi}}\exp\left[ -\frac{y^2}{2} \right] dy.
\end{IEEEeqnarray*}
Since $W_1 >0$ can be arbitrarily large, we have
%\begin{multline*}
%\lefteqn{\lim_{n \to \infty} \Pr\left\{ \frac{-\!\log P_{Y^n_{2}}(Y^n_{2}) - nH(Y_2)}{\sqrt{n}\sigma_2} \right. }  \\
%\left. >  \frac{ n\left(H(Y_1)-H(Y_2)\right)}{\sqrt{n}\sigma_2} +\frac{ T_2 - 2 \gamma }{\sigma_2} \right\} = 0.
%\end{multline*}
\begin{equation*}
\lim_{n \to \infty} \Pr\left\{ \frac{-\!\log P_{Y^n_{2}}(Y^n_{2}) - nH(Y_2)}{\sqrt{n}\sigma_2} >  \frac{ n\left(H(Y_1)-H(Y_2)\right)}{\sqrt{n}\sigma_2} +\frac{ T_2 - 2 \gamma }{\sigma_2} \right\} = 0.
\end{equation*}
Thus, from the asymptotic normality it follows that
\begin{align*} 
\sum_{i=1}^2 & \limsup_{n \to \infty} \Pr\left\{ \frac{\!-\!\log P_{Y_i^n}(Y_i^n)\!-\!nH(Y_1)}{\sqrt{n}}\!>\!T_2\!-\!2 \gamma  \right\}w(i)  \nonumber \\
 = & \limsup_{n \to \infty} \Pr\left\{ \frac{\!-\!\log P_{Y_1^n}(Y_1^n)\!-\!nH(Y_1)}{\sqrt{n} \sigma_1}\!>\!\frac{T_2\!-\!2 \gamma}{\sigma_1}  \right\} w(1) \\
 = & w(1) \int^{\infty}_{\frac{T_2 - 2 \gamma}{\sigma_1}} \frac{1}{\sqrt{2\pi}}\exp\left[ -\frac{z^2}{2} \right] dz \\
 \to & w(1) \int^{\infty}_{\frac{{T}_2}{\sigma_1}} \frac{1}{\sqrt{2\pi}}\exp\left[ -\frac{z^2}{2} \right] dz = \frac{\delta}{2},
\end{align*} 
by letting $\gamma \to 0$.
Thus, we have proved the direct part of {Case II}.

\vspace*{\baselineskip}

b) In {\it Case III}: we have%, let $\overline{T}_3 = T_3 - 2 \gamma$, where $\gamma$ is an arbitrary small number, similarly we have
\begin{align*} 
\lefteqn{\limsup_{n \to \infty} \frac{1}{2} d \left( \phi_n(U_{M_n}),Y^n \right)} \nonumber \\ 
& \leq \sum_{i=1}^2 \limsup_{n \to \infty} \Pr\left\{ \frac{-\!\log P_{Y_i^n}(Y_i^n)\!-\!nH(Y_2)}{\sqrt{n}}\!>\!{T}_3\!-\! 2\gamma \right\}w(i) \nonumber \\
%& \leq  w(1) \\
%& \ \ \   +\!\limsup_{n \to \infty} \Pr\left\{ \frac{\!-\!\log P_{Y_2^n}(Y_2^n)\!-\!nH(Y_2)}{\sqrt{n} \sigma_2} \!>\!\frac{{T}_3\!-\!2\gamma}{\sigma_2}  \right\} w(2) \\
& \leq  w(1)  + \limsup_{n \to \infty} \Pr\left\{ \frac{\!-\!\log P_{Y_2^n}(Y_2^n)\!-\!nH(Y_2)}{\sqrt{n} \sigma_2} \!>\!\frac{{T}_3\!-\!2\gamma}{\sigma_2}  \right\} w(2) \\& \to w(1)+ w(2) \int^{\infty}_{\frac{{T}_3}{\sigma_2}} \frac{1}{\sqrt{2\pi}}\exp\left[ -\frac{z^2}{2} \right] dz
\  = \frac{\delta}{2}
\end{align*} 
by letting $\gamma \to 0$. Consequently, similarly to the proof of {Case I} or {Case II}, we can prove the direct part of {Case III}.

\vspace*{\baselineskip}

{\it 2) Converse Part}:

a) In {\it Case II}: we consider a constant $T_2' < T_2$ and we assume that $T_2'$ is $(H(Y_1),\delta)$-achievable. Notice that there exists a constant $\gamma >0$ satisfying $T_2' + \gamma < T_2$.
Then, similarly to the proof of {Case I}, we have
\begin{align*} 
\lefteqn{\liminf_{n \to \infty} \frac{1}{2} d(\phi_n(U_{M_n}),Y^n)} \\
%& \geq \sum_{i=1}^2  \liminf_{n \to \infty} \Pr\left\{\frac{\!-\! \log P_{Y_i^n}(Y_i^n)\!-\!nH(Y_1)}{\sqrt{n}}\!>\!T_2'\!+\!2 \gamma\!+\!\gamma_n \right\} \\
%& \ \ \ \cdot w(i) \nonumber \\
& \geq \sum_{i=1}^2  \liminf_{n \to \infty} \Pr\left\{\frac{\!-\! \log P_{Y_i^n}(Y_i^n)\!-\!nH(Y_1)}{\sqrt{n}}\!>\!T_2'\!+\!2 \gamma\!+\!\gamma_n \right\} w(i) \nonumber \\
& \geq   \liminf_{n \to \infty} \Pr\left\{\frac{- \log P_{Y_1^n}(Y_1^n) - nH(Y_1)}{\sqrt{n}} > T_2'\!+\!3\gamma \right\} w(1) \nonumber \\
& = \ w(1) \int^{\infty}_{\frac{T'_2+3\gamma}{\sigma_1}} \frac{1}{\sqrt{2\pi}}\exp\left[ -\frac{z^2}{2} \right] dz \\
& >  \  w(1) \int^{\infty}_{\frac{T_2}{\sigma_1}} \frac{1}{\sqrt{2\pi}}\exp\left[ -\frac{z^2}{2} \right] dz = \frac{\delta}{2},
\end{align*}
because we can let $T_2' + 3\gamma < T_2$ by letting $\gamma \to 0$.
Hence, we have shown the converse part by using the similar argument to {Case I}.

b) Similarly, in {\it Case III}: we consider a constant $T_3' < T_3$ and we assume that $T_3'$ is $(H(Y_2),\delta)$-achievable. Notice that there exists a constant $\gamma >0$ satisfying $T_3' + \gamma < T_3$. %Let $\overline{T}_3$ be $\overline{T}_3 \triangleq T_3' + \gamma$ for short.
Then, we have
\begin{align} \label{eq:4-3-c-1}
\lefteqn{\liminf_{n \to \infty} \frac{1}{2} d(\phi_n(U_{M_n}),Y^n)} \nonumber \\
%& \geq \sum_{i=1}^2 \liminf_{n \to \infty} \Pr\left\{ \frac{\!-\!\log P_{Y_i^n}(Y_i^n)\!-\!nH(Y_2)}{\sqrt{n}}\!>\!{T}'_3\!+\! 2\gamma\!+\!\gamma_n  \right\} \nonumber \\
%& \ \ \  \cdot w(i) \nonumber \\
& \geq \sum_{i=1}^2 \liminf_{n \to \infty} \Pr\left\{ \frac{\!-\!\log P_{Y_i^n}(Y_i^n)\!-\!nH(Y_2)}{\sqrt{n}}\!>\!{T}'_3\!+\! 2\gamma\!+\!\gamma_n  \right\} w(i) \nonumber \\
%& \geq \liminf_{n \to \infty} \Pr\left\{ \frac{\!-\!\log P_{Y_1^n}(Y_1^n)\!-\!nH(Y_1)}{\sqrt{n} \sigma_1} \right. \nonumber \\
%& \ \ \ \left. > \frac{n\left( H(Y_2)- H(Y_1)\right) }{\sqrt{n} \sigma_1} + \frac{{T}'_3 + 3 \gamma}{\sigma_1}  \right\} w(1) \nonumber \\
& \geq \liminf_{n \to \infty} \Pr\left\{ \frac{\!-\!\log P_{Y_1^n}(Y_1^n)\!-\!nH(Y_1)}{\sqrt{n} \sigma_1} > \frac{n\left( H(Y_2)- H(Y_1)\right) }{\sqrt{n} \sigma_1} + \frac{{T}'_3 + 3 \gamma}{\sigma_1}  \right\} w(1) \nonumber \\
& \ \ \ + \liminf_{n \to \infty} \Pr\left\{ \frac{\!-\!\log P_{Y_2^n}(Y_2^n)\!-\!nH(Y_2)}{\sqrt{n} \sigma_2}\!>\!\frac{{T}'_3\!+\!3\gamma}{\sigma_2}  \right\} w(2).
%& + \limsup_{n \to \infty} \Pr\left\{ \frac{\!-\!\log P_{Y_2^n}(Y_2^n)\!-\!nH(Y_2)}{\sqrt{n} \sigma_2} > \frac{\overline{T}_3}{\sigma_2}  \right\} w(2).
\end{align} 
Then, the first term of the right-hand side of the above inequality can be determined as follows.
Notice that $H(Y_1) > H(Y_2)$ holds. 
This means that for any constant $W_1 > 0$
\[
\frac{n\left( H(Y_2)- H(Y_1)\right) }{\sqrt{n} \sigma_1} + \frac{{T}'_3 + 3\gamma}{\sigma_1} < - W_1,
\]
holds for sufficiently large $n$. Thus, %taking account of $H(Y_1) > H(Y_2)$, we have
\begin{align*}
%\lefteqn{\liminf_{n \to \infty} \Pr\left\{ \frac{\!-\!\log P_{Y_1^n}(Y_1^n)\!-\!nH(Y_1)}{\sqrt{n} \sigma_1} \right.} \\
%& \ \ \ \left. > \frac{n\left( H(Y_2)- H(Y_1)\right) }{\sqrt{n} \sigma_1} + \frac{{T}'_3 + 3\gamma}{\sigma_1}  \right\} \\
\lefteqn{\liminf_{n \to \infty} \Pr\left\{ \frac{\!-\!\log P_{Y_1^n}(Y_1^n)\!-\!nH(Y_1)}{\sqrt{n} \sigma_1}  > \frac{n\left( H(Y_2)- H(Y_1)\right) }{\sqrt{n} \sigma_1} + \frac{{T}'_3 + 3\gamma}{\sigma_1}  \right\} }\\
& \geq  \liminf_{n \to \infty} \Pr\left\{ \frac{-\!\log P_{Y^n_{1}}(Y^n_{1}) - nH(Y_1)}{\sqrt{n} \sigma_1} > - W_1 \right\} \\
& =  \int_{-W_1}^{\infty} \frac{1}{\sqrt{2\pi}}\exp\left[ -\frac{y^2}{2} \right] dy,
\end{align*}
holds. 
Since $W_1 >0$ can be arbitrarily large, we have
%\begin{multline*}
%\lefteqn{\lim_{n \to \infty} \Pr\left\{ \frac{-\!\log P_{Y^n_{1}}(Y^n_{1}) - nH(Y_1)}{\sqrt{n}\sigma_1} \right. }  \\
%\left. >  \frac{ n\left(H(Y_2)-H(Y_1)\right)}{\sqrt{n}\sigma_1} +\frac{ {T}'_3 + 3 \gamma }{\sigma_1} \right\} = 1.
%\end{multline*}
\begin{equation*}
\lim_{n \to \infty} \Pr\left\{ \frac{-\!\log P_{Y^n_{1}}(Y^n_{1}) - nH(Y_1)}{\sqrt{n}\sigma_1}  >  \frac{ n\left(H(Y_2)-H(Y_1)\right)}{\sqrt{n}\sigma_1} +\frac{ {T}'_3 + 3 \gamma }{\sigma_1} \right\} = 1.
\end{equation*}
Substituting the above equality into (\ref{eq:4-3-c-1}),  by virtue of the asymptotic normality, it holds that
\begin{align*}
\lefteqn{\liminf_{n \to \infty} \frac{1}{2} d(\phi_n(U_{M_n}),Y^n)} \\
& \geq \sum_{i=1}^2 \liminf_{n \to \infty} \Pr\left\{ \frac{\!-\!\log P_{Y_i^n}(Y_i^n)\!-\!nH(Y_2)}{\sqrt{n}} >{T}'_3\!+\!3 \gamma  \right\}w(i) \nonumber \\
%& = \  w(1) \nonumber \\
%& \ \ \ \!+\! \liminf_{n \to \infty} \Pr\left\{ \frac{\!-\!\log P_{Y_2^n}(Y_2^n)\!-\!nH(Y_2)}{\sqrt{n} \sigma_2}\!>\!\frac{{T}'_3\!+\!3 \gamma}{\sigma_2}  \right\} w(2) \\
& = \  w(1) + \liminf_{n \to \infty} \Pr\left\{ \frac{\!-\!\log P_{Y_2^n}(Y_2^n)\!-\!nH(Y_2)}{\sqrt{n} \sigma_2}\!>\!\frac{{T}'_3\!+\!3 \gamma}{\sigma_2}  \right\} w(2) \\
& = \  w(1) + w(2) \int^{\infty}_{\frac{T'_3+3 \gamma}{\sigma_1}} \frac{1}{\sqrt{2\pi}}\exp\left[ -\frac{z^2}{2} \right] dz \\
& >  \ w(1) + w(2) \int^{\infty}_{\frac{T_3}{\sigma_1}} \frac{1}{\sqrt{2\pi}}\exp\left[ -\frac{z^2}{2} \right] dz = \frac{\delta}{2},
\end{align*} 
because we can let $T_3' + 3\gamma < T_3$ by letting $\gamma \to 0$.
Hence, we have shown the proofs similarly to {Case I}.
\end{IEEEproof}
%
% appendix d
%
\section{Proof of theorem \ref{theo2}} \label{app3}
\renewcommand{\theequation}{D.\arabic{equation}}
\setcounter{equation}{0}
{\it 1) Direct Part}: 

We prove that $L_1 = T_7 + \gamma$ is an $(H(Y_1),\varepsilon)$-achievable, where $\gamma >0$ is an arbitrary small constant.
Set $M_n = e^{nH(Y_1) + \sqrt{n}L_1}$. Then, obviously we have
$$
\limsup_{n \rightarrow \infty}  \frac{\log M_n - n H(Y_1) }{\sqrt{n}}  \leq L_1.
$$
Thus, it is enough to show that  there exists an $(n, M_n, \varepsilon_n)$ code satisfying 
$$
\limsup_{n \rightarrow \infty} \varepsilon_n \leq \varepsilon.
$$
From Lemma \ref{lemma3}, there exists an $(n, M_n, \epsilon_n)$ code that satisfies
\begin{eqnarray} \label{5-1}
\varepsilon_n & \leq &\Pr \left\{ \frac{1}{\sqrt{n}} \log\frac{1}{P_{Y^n}(Y^n)} \geq \frac{1}{\sqrt{n}} \log M_n \right\} \nonumber \\
& = & \Pr \left\{ \frac{1}{\sqrt{n}} \log\frac{1}{P_{Y^n}(Y^n)} \geq \frac{nH(Y_1)+ \sqrt{n}L_1}{\sqrt{n}}  \right\} \nonumber \\
& =& \Pr \left\{ \frac{-\log P_{Y^n}(Y^n) - nH(Y_1)}{\sqrt{n}} \geq L_1  \right\} \nonumber \\
& = & \sum_{i=1}^2\Pr \left\{ \frac{-\log P_{Y^n}(Y_i^n) - nH(Y_1)}{\sqrt{n}} \geq L_1  \right\} w(i). 
\end{eqnarray}
The last equality is derived from the definition of the mixed source. Then, from Lemma \ref{lemma}, we have
\begin{eqnarray*}
\lefteqn{\limsup_{n \to \infty} \sum_{i=1,2} \Pr \left\{ \frac{-\log P_{Y^n}(Y_i^n) - nH(Y_1)}{\sqrt{n}} \geq L_1  \right\}w(i)} \nonumber \\
%&\!\leq\!&\sum_{i=1}^2 \limsup_{n \to \infty} \Pr \left\{ \frac{\!-\!\log P_{Y_i^n}(Y_i^n)\!-\!nH(Y_1)}{\sqrt{n}} \geq L_1\!-\!\gamma_n  \right\} \nonumber \\
%&  & \cdot w(i) \nonumber \\
&\!\leq\!&\sum_{i=1}^2 \limsup_{n \to \infty} \Pr \left\{ \frac{\!-\!\log P_{Y_i^n}(Y_i^n)\!-\!nH(Y_1)}{\sqrt{n}} \geq L_1\!-\!\gamma_n  \right\} w(i) \nonumber \\
%&\!\leq\!&\sum_{i=1}^2 \limsup_{n \to \infty} \Pr \left\{ \frac{\!-\!\log P_{Y_i^n}(Y_i^n)\!-\!nH(Y_1)}{\sqrt{n}\sigma_i} \geq \frac{L_1\!-\!\gamma}{\sigma_i}  \right\} \nonumber \\
%&  & \cdot w(i) \\
&\!\leq\!&\sum_{i=1}^2 \limsup_{n \to \infty} \Pr \left\{ \frac{\!-\!\log P_{Y_i^n}(Y_i^n)\!-\!nH(Y_1)}{\sqrt{n}\sigma_i} \geq \frac{L_1\!-\!\gamma}{\sigma_i}  \right\}  w(i) \\
&\!=\!&\sum_{i=1}^2 \limsup_{n \to \infty} \Pr \left\{ \frac{\!-\!\log P_{Y_i^n}(Y_i^n)\!-\!nH(Y_1)}{\sqrt{n}\sigma_i} \geq \frac{T_7}{\sigma_i}  \right\} w(i),
\end{eqnarray*}
because $\gamma_n$ is specified in Lemma \ref{lemma}, and so $\gamma > \gamma_n$ holds for sufficiently large $n$.

Then, noting that $H(Y_1)=H(Y_2)$ holds,  from the asymptotic normality, we have
\begin{align*}
%&\limsup_{n \rightarrow \infty}\Pr\left\{  \frac{\!-\log P_{Y_i^n}(Y_i^n)- nH(Y_1) }{\sqrt{n} \sigma_i} \geq \frac{T_7}{\sigma_i} \right\} \\
%& = \int^{\infty}_{\frac{T_4}{\sigma_i}} \frac{1}{\sqrt{2\pi}}\exp\left[ -\frac{z^2}{2} \right] dz,
\limsup_{n \rightarrow \infty}\Pr\left\{  \frac{\!-\log P_{Y_i^n}(Y_i^n)- nH(Y_1) }{\sqrt{n} \sigma_i} \geq \frac{T_7}{\sigma_i} \right\}  = \int^{\infty}_{\frac{T_7}{\sigma_i}} \frac{1}{\sqrt{2\pi}}\exp\left[ -\frac{z^2}{2} \right] dz,
\end{align*} 
for $i=1,2$.
Thus, we have
\begin{align*}
%&\limsup_{n \to \infty} \sum_{i=1}^2 \Pr \left\{ \frac{-\log P_{Y^n}(Y_i^n) - nH(Y_1)}{\sqrt{n}} \geq L_1  \right\}w(i) \nonumber \\
%& \leq  \sum_{i=1}^2 w(i) \int^{\infty}_{\frac{T_7}{\sigma_i}} \frac{1}{\sqrt{2\pi}}\exp\left[ -\frac{z^2}{2} \right] dz.
\limsup_{n \to \infty} \sum_{i=1}^2 \Pr \left\{ \frac{-\log P_{Y^n}(Y_i^n) - nH(Y_1)}{\sqrt{n}} \geq L_1  \right\}w(i)  \leq  \sum_{i=1}^2 w(i) \int^{\infty}_{\frac{T_7}{\sigma_i}} \frac{1}{\sqrt{2\pi}}\exp\left[ -\frac{z^2}{2} \right] dz.
\end{align*}
Substituting the above inequality into (\ref{5-1}), we have
\begin{eqnarray*}
\limsup_{n \to \infty}\epsilon_n  \leq  \sum_{i=1}^2 w(i) \int^{\infty}_{\frac{T_7}{\sigma_i}} \frac{1}{\sqrt{2\pi}}\exp\left[\!-\!\frac{z^2}{2} \right] dz = \varepsilon,
\end{eqnarray*}
where the last equality is derived from the definition of $T_7$ given by (\ref{T4}).
%Finally, notice that for any numerical sequence $\{ a_n \}_{n=1}^\infty$ satisfying $0 \leq a_n <1 $$(n = 1,2,\cdots)$ and any constant $A$, if we assume that
Therefore, Direct Part has been proved.

\vspace*{\baselineskip}

{\it 2) Converse Part}:

Let us assume that $L_2$ satisfying $L_2 < T_7$ is $(H(Y_1),\varepsilon)$-achievable. Then we shall show a contradiction. Notice that there exists a constant $\gamma >0$ such that $L_2 + 3\gamma  < T_7$ holds.

Then from the assumption there must exist an $(n, M_n, \epsilon_n)$ code such that 
\[
\limsup_{n \rightarrow \infty} \varepsilon_n \leq \varepsilon,
\]
\[
\limsup_{n \rightarrow \infty}  \frac{\log M_n - n H(Y_1) }{\sqrt{n}}  \leq L_2,
\]
This means that for sufficiently large $n$, there must exist an $(n, M_n, \epsilon_n)$ code such that
\begin{eqnarray} \label{eq:5-c--3}
\frac{\log M_n }{\sqrt{n}} < \frac{nH(Y_1)}{\sqrt{n}} + L_2 +\gamma.
\end{eqnarray}
for any $\gamma >0$.
On the other hand, from Lemma \ref{lemma4}, any $(n, M_n, \epsilon_n)$ code satisfies
\begin{eqnarray*}
\varepsilon_n &\!\geq\!& \Pr \left\{ \frac{1}{\sqrt{n}} \log\frac{1}{P_{Y^n}(Y^n)} \geq \frac{1}{\sqrt{n}} \log M_n\!+\!\gamma \right\}\!-\!e^{-\sqrt{n}\gamma}.
\end{eqnarray*}
Thus, from (\ref{eq:5-c--3}) it follows that
\begin{eqnarray*}
\varepsilon_n &\!\geq\!& \Pr \left\{ \frac{1}{\sqrt{n}} \log\frac{1}{P_{Y^n}(Y^n)} \geq \frac{1}{\sqrt{n}} \log M_n\!+\!\gamma \right\}\!-\!e^{-\sqrt{n}\gamma} \\
%& \geq & \Pr \left\{ \frac{1}{\sqrt{n}} \log\frac{1}{P_{Y^n}(Y^n)} \geq \frac{nH(Y_1)}{\sqrt{n}} + L_2 + 2\gamma \right\} \\
%&        & \!-\!e^{-\sqrt{n}\gamma} \\
& \geq & \Pr \left\{ \frac{1}{\sqrt{n}} \log\frac{1}{P_{Y^n}(Y^n)} \geq \frac{nH(Y_1)}{\sqrt{n}} + L_2 + 2\gamma \right\} - e^{-\sqrt{n}\gamma} \\
%& \geq & \Pr \left\{ \frac{- \log P_{Y^n}(Y^n) - nH(Y_1)}{\sqrt{n}}  \geq L_2 + 2\gamma \right\} \\
%&        & \!-\!e^{-\sqrt{n}\gamma} \\
%& \geq & \Pr \left\{ \frac{- \log P_{Y^n}(Y^n) - nH(Y_1)}{\sqrt{n}}  \geq L_2 + 2\gamma \right\} - e^{-\sqrt{n}\gamma} \\
%& = & \sum_{i=1}^2\Pr \left\{ \frac{\!-\!\log P_{Y^n}(Y_i^n)\!-\!nH(Y_1)}{\sqrt{n}}  \geq L_2\!+\!2\gamma \right\} w(i) \\
%&        & \!-\!e^{-\sqrt{n}\gamma},
& = & \sum_{i=1}^2\Pr \left\{ \frac{\!-\!\log P_{Y^n}(Y_i^n)\!-\!nH(Y_1)}{\sqrt{n}}  \geq L_2\!+\!2\gamma \right\} w(i) - e^{-\sqrt{n}\gamma},\end{eqnarray*}
for sufficiently large $n$, where the last equality is derived from the definition of the mixed source.
From Lemma \ref{lemma}, %any $(n, M_n, \epsilon_n)$ code satisfies
we have
\begin{eqnarray} \label{5-3}
%\lefteqn{\liminf_{n \to \infty} \varepsilon_n} \nonumber \\
%&\!\geq\!& \liminf_{n \to \infty} \sum_{i=1}^2\Pr \left\{ \frac{\!-\!\log P_{Y^n}(Y_i^n)\!-\!nH(Y_1)}{\sqrt{n}}  \geq L_2\!+\!2\gamma \right\} \nonumber \\
%&    & \cdot  w(i) \nonumber \\
%&\!\geq\!& \sum_{i=1}^2 \liminf_{n \to \infty}\Pr \left\{ \frac{\!-\!\log P_{Y_i^n}(Y_i^n)\!-\!nH(Y_1)}{\sqrt{n}}  \geq L_2 +2\gamma \right. \nonumber \\
%&    & \ \ \ \ +\gamma_n \Biggr\}   w(i) \nonumber \\
%&\!\geq\!& \sum_{i=1}^2 \liminf_{n \to \infty}\Pr \left\{ \frac{\!-\!\log P_{Y_i^n}(Y_i^n)\!-\!nH(Y_1)}{\sqrt{n}}  \geq L_2\!+\!3\gamma \right\} \nonumber \\
%&    & \cdot  w(i),
\liminf_{n \to \infty} \varepsilon_n & \geq & \liminf_{n \to \infty} \sum_{i=1}^2\Pr \left\{ \frac{\!-\!\log P_{Y^n}(Y_i^n)\!-\!nH(Y_1)}{\sqrt{n}}  \geq L_2\!+\!2\gamma \right\} w(i) \nonumber \\
&\!\geq\!& \sum_{i=1}^2 \liminf_{n \to \infty}\Pr \left\{ \frac{\!-\!\log P_{Y_i^n}(Y_i^n)\!-\!nH(Y_1)}{\sqrt{n}}  \geq L_2 +2\gamma +\gamma_n \right\}   w(i) \nonumber \\
&\!\geq\!& \sum_{i=1}^2 \liminf_{n \to \infty}\Pr \left\{ \frac{\!-\!\log P_{Y_i^n}(Y_i^n)\!-\!nH(Y_1)}{\sqrt{n}}  \geq L_2\!+\!3\gamma \right\} w(i),
\end{eqnarray}
because $\gamma_n$ is specified in Lemma \ref{lemma} and so $\gamma_n < \gamma$ holds for sufficiently large $n$.
%Let $L_2'$ be as $L_2' = L_2 + 3\gamma$ for short. 
Then, by virtue of the asymptotic normality we have
\begin{align*}
\lefteqn{\sum_{i=1}^2 \liminf_{n \to \infty}\Pr \left\{ \frac{\!-\!\log P_{Y_i^n}(Y_i^n)\!-\!nH(Y_1)}{\sqrt{n}}  \geq L_2 + 3\gamma \right\} w(i)} \\
&\!=\! \sum_{i=1}^2 \liminf_{n \to \infty}\Pr \left\{ \frac{\!-\!\log P_{Y_i^n}(Y_i^n)\!-\!nH(Y_1)}{\sqrt{n}\sigma_i} \!\geq\!\frac{L_2\!+\!3\gamma}{\sigma_i} \right\} w(i) \\
&\!=\! \sum_{i=1}^2 w(i) \int^{\infty}_{\frac{L_2 + 3\gamma}{\sigma_i}} \frac{1}{\sqrt{2\pi}}\exp\left[ -\frac{z^2}{2} \right] dz \\
&\!>\! \sum_{i=1}^2 w(i) \int^{\infty}_{\frac{T_7}{\sigma_i}} \frac{1}{\sqrt{2\pi}}\exp\left[ -\frac{z^2}{2} \right] dz = \varepsilon,
\end{align*}
because we can let $L_2 + 3\gamma < T_7$ holds by letting $\gamma \to 0$ and  the last equality is from (\ref{T4}).
Thus, substituting the above into (\ref{5-3}), it must hold that
\begin{eqnarray*}
\liminf_{n \to \infty} \varepsilon_n > \varepsilon.
\end{eqnarray*}
This is a contradiction and we conclude that $L_2$ satisfying $L_2 < T_7$ is not an $(H(Y_1),\varepsilon)$-achievable.
\IEEEQED

\vspace*{\baselineskip}

\begin{IEEEproof}[Proofs of Case II and Case III]

The proofs of Case II and Case III can be shown by the same arguments as the proof of Case I of Theorem \ref{theo2}.
\end{IEEEproof}

% use section* for acknowledgement
\section*{Acknowledgment}
The authors are very grateful to the anonymous reviewers for their helpful comments, which have occasioned to improve the main results so as to be much more general and transparent.
%The authors would like to thank...
The first author is very grateful to Prof. Toshiyasu Matsushima of Waseda University for his valuable discussions and comments.
% Can use something like this to put references on a page
% by themselves when using endfloat and the captionsoff option.
\ifCLASSOPTIONcaptionsoff
  \newpage
\fi

\end{document}